\documentclass[11pt]{article}

\PassOptionsToPackage{table,xcdraw,dvipsnames}{xcolor}

\usepackage[final]{acl}

\usepackage{times}
\usepackage{latexsym}

\usepackage[T1]{fontenc}

\usepackage[utf8]{inputenc}

\usepackage{microtype}

\usepackage{inconsolata}

\usepackage{graphicx}

\usepackage{tikz}
\usetikzlibrary{tikzmark, positioning, shapes.geometric, arrows.meta, fit, backgrounds, calc}

\usepackage{amsmath}
\usepackage{amsfonts}

\usepackage{booktabs}
\usepackage{makecell}
\usepackage{multirow}

\usepackage{xspace}

\newcommand{\numattacks}{19\xspace}
\newcommand{\numdefenses}{15\xspace}

\newcommand{\resultU}{\overline{\%\mkern-4mu\uparrow\mkern-4mu U}}
\newcommand{\resultASR}{\overline{\%\mkern-4mu\downarrow\mkern-4mu ASR}}
\newcommand{\resultParam}{|\Theta|}
\newcommand{\resultT}{\Delta T}

\title{CASCADE Against Jailbreaks: \\ Combination Across Stages with \\ Controlled Attack-Defense Evaluation}

\author{Jiale Luo \\
  School of Computing \\
  National University of Singapore \\
\texttt{luojiale@u.nus.edu} \\\And
  Eric Han \\
  School of Computing \\
  National University of Singapore \\
\texttt{eric\_han@nus.edu.sg} \\}

\providecommand{\attackColor}{red!75!black}
\providecommand{\attackFill}{red!12}
\providecommand{\defenseColor}{blue!75!black}
\providecommand{\defenseFill}{blue!12}
\providecommand{\neutralColor}{gray!75!black}
\providecommand{\llmColor}{black!70}
\providecommand{\llmFill}{gray!10}
\providecommand{\arrowColor}{black!85}
\providecommand{\panelColor}{black}

\providecommand{\chevAspectH}{0.406}
\providecommand{\chevAspectN}{0.286}

\newcommand{\drawChevron}[3]{\draw[#1]
        ($(#2)+(-#3,{#3*\chevAspectH})$) --
        ($(#2)+(0,{#3*\chevAspectN})$) --
        ($(#2)+(#3,{#3*\chevAspectH})$) --
        ($(#2)+(#3,{-#3*\chevAspectN})$) --
        ($(#2)+(0,{-#3*\chevAspectH})$) --
        ($(#2)+(-#3,{-#3*\chevAspectN})$) -- cycle;
}

\tikzset{
    lab/.style={
        font=\rmfamily\scriptsize,
        align=center
    },
    boldlab/.style={
        font=\rmfamily\scriptsize\bfseries,
        align=center
    },
    pentagonBase/.style={
        regular polygon,
        regular polygon sides=5,
        thick,
        minimum size=5mm,
        inner sep=-1.5pt,
        font=\rmfamily\tiny\bfseries,
        align=center
    }
} 
\begin{document}
\maketitle
\begin{abstract}
Defenses against jailbreak attacks on Large Language Models (LLMs) operate at different pipeline stages, such as input modification or output guard, but it remains unclear which defenses to deploy at each stage and how to combine them. Prior empirical studies, fragmented by inconsistent attack-success-rate definitions and experimental settings, have evaluated defenses largely in isolation. Here we present the first systematic study, to our knowledge, of defense combinations both within and across pipeline stages, under a consistent threat model of direct, black-box, single-turn attacks. Our decision framework standardizes evaluation through a principled attack-success-rate formulation with controlled query budgets, together with explicit fairness rules. Across \numattacks attacks and \numdefenses defenses, we find that no single defense is universally best, but well-chosen combinations achieve substantial safety with minimal utility degradation, yielding practical recommendations for layered defense pipelines.
\end{abstract}

\section{Introduction}

Large Language Models (LLMs) such as GPT, Claude, and LLaMA underpin applications ranging from chatbots and code generation to automated scoring of student responses (\citealp{openai2024gpt4technicalreport}; \citealp{LEE2024100213}). Advances such as ReAct-style tool use \citep{DBLP:conf/iclr/YaoZYDSN023} and personalized AI assistants \citep{steinberger2026openclaw} extend LLM autonomy to external actions, including code execution, shell commands, and other consequential operations.

This growing autonomy expands the attack surface. \citet{owasp_llm_top10_2025} identifies prompt injection as the top threat, embedding adversarial input in user prompts to manipulate LLMs \citep{perez2022ignorepreviouspromptattack}. Jailbreak, a prevalent form of prompt injection \citep{owasp_llm_top10_2025}, explicitly aims to circumvent safety measures, with consequences ranging from harmful generations (e.g., self-harm guidance) to unauthorized command execution \citep{vassilev2025adversarial}.

\providecommand{\elemPitch}{0.55}
\providecommand{\elemCenter}{0.05}
\providecommand{\captionOffset}{-0.05}
\providecommand{\pentaShift}{-0.035}
\providecommand{\panelPitch}{3.10}

\providecommand{\chevHalfW}{0.65}
\providecommand{\chevNudge}{0.2pt}

\providecommand{\panelWidth}{8.8}
\providecommand{\arrowGap}{0.10}
\providecommand{\titleInset}{0.12}
\providecommand{\titleDrop}{-0.28}
\providecommand{\captionInset}{0.10}
\providecommand{\chevStackPitch}{0.45}

\providecommand{\chevAt}[2][]{\drawChevron{chevStage}{#2}{\chevHalfW}\node[lab] at (#2) {#1};
}
\providecommand{\chevDropAt}[2][]{\drawChevron{chevDrop}{#2}{\chevHalfW}\node[lab] at (#2) {#1};
}
\providecommand{\chevAtNoFill}[2][]{\drawChevron{chevStage, fill=none}{#2}{\chevHalfW}\node[lab] at (#2) {#1};
}
\providecommand{\chevLLMAt}[1]{\drawChevron{chevLLM}{#1}{\chevHalfW}\node[boldlab] at (#1) {LLM};
}

\newcommand{\chevStackAt}[2]{\coordinate (#11c) at ($(#2)+(0,\elemCenter+\chevStackPitch)+(0,-\chevNudge)$);
    \coordinate (#12c) at ($(#2)+(0,\elemCenter)+(0,-\chevNudge)$);
    \coordinate (#13c) at ($(#2)+(0,\elemCenter-\chevStackPitch)+(0,-\chevNudge)$);
}

\begin{figure}[t]
\centering
\resizebox{\columnwidth}{!}{
\begin{tikzpicture}[
    font=\rmfamily\footnotesize,
    >=Stealth,
    rowPanel/.style={
        draw=none,
        fill=none,
        minimum width=\panelWidth cm,
        minimum height=\panelPitch cm,
        inner sep=0pt,
        outer sep=0pt
    },
    outerPanel/.style={
        draw=\panelColor,
        fill=none,
        rounded corners=2pt,
        thick
    },
    hsep/.style={
        draw=\panelColor,
        thick
    },
    title/.style={
        font=\rmfamily\footnotesize\bfseries,
        anchor=west
    },
    caption/.style={
        font=\rmfamily\scriptsize,
        align=center,
        text depth=0.5ex
    },
colCap/.style={caption, minimum width=1.30cm},
    wideCap/.style={caption, minimum width=1.60cm},
    arrowlab/.style={
        font=\rmfamily\scriptsize,
        align=center,
        fill=none,
        inner sep=1pt
    },
    box/.style={
        rectangle,
        thick,
        inner ysep=0pt,
        inner xsep=4pt,
        minimum width=1.30cm,
        minimum height=0.45cm,
        font=\rmfamily\scriptsize
    },
    subcat/.style={box, draw=\neutralColor},
    subcatSel/.style={box, draw=\attackColor},
    stage/.style={box, draw=\neutralColor},
    stageSel/.style={box, draw=\defenseColor},
    chevStage/.style={
        draw=\defenseColor,
        fill=\defenseFill,
        thick,
        line join=miter
    },
    chevDrop/.style={
        draw=\neutralColor,
        fill=none,
        dash pattern=on 2.8pt off 1pt,
        line join=miter,
        thick
    },
    chevLLM/.style={
        draw=\llmColor,
        fill=\llmFill,
        thick,
        line join=miter
    },
    atk/.style={pentagonBase, draw=\attackColor},
    atkSel/.style={atk, fill=\attackFill},
    atkDrop/.style={
        pentagonBase,
        draw=\neutralColor,
        dash pattern=on 2.8pt off 1pt,
        line join=round
    },
    circleBase/.style={
        circle,
        thick,
        minimum size=4.5mm,
        inner sep=0pt,
        font=\rmfamily\scriptsize\bfseries
    },
    def/.style={circleBase, draw=\defenseColor},
    defSel/.style={def, fill=\defenseFill},
    defDrop/.style={
        circleBase,
        draw=\neutralColor,
        dash pattern=on 2.8pt off 1pt,
        line join=round
    },
    llm/.style={
        rectangle,
        draw=\llmColor,
        fill=\llmFill,
        thick,
        rounded corners=1pt,
        inner ysep=0pt,
        inner xsep=4pt,
text height=1.5ex,
        text depth=0.4ex,
        minimum width=0.80cm,
        minimum height=0.45cm,
        align=center,
        font=\rmfamily\scriptsize\bfseries
    },
    arr/.style={->, thick, draw=\arrowColor},
    zoom/.style={dashed, thick, draw=\neutralColor}
]

\pgfmathsetmacro{\elemTop}{\elemCenter + \elemPitch}
\pgfmathsetmacro{\elemBot}{\elemCenter - \elemPitch}

\node[rowPanel] (p1) at (0,0) {};
\node[rowPanel] (p2) at (0,{-\panelPitch}) {};
\node[rowPanel] (p3) at (0,{-2*\panelPitch}) {};
\node[rowPanel] (p4) at (0,{-3*\panelPitch}) {};

\begin{scope}[on background layer]
    \draw[outerPanel] (p1.north west) rectangle (p4.south east);

    \draw[hsep] (p1.south west) -- (p1.south east);
    \draw[hsep] (p2.south west) -- (p2.south east);
    \draw[hsep] (p3.south west) -- (p3.south east);
\end{scope}

\node[title] at ($(p1.north west)+(\titleInset,\titleDrop)$)
    {(P1) Initial selection of attacks};

\node[caption,anchor=south west] (e1descsubcat) at ($(p1.south west)+(\captionInset,\captionOffset)$)
    {subcategories of\\similar attacks};
\node[caption,anchor=south east] (e1descrep) at ($(p1.south east)+(-\captionInset,\captionOffset)$)
    {per-cat. attack\\representatives};

\coordinate (e1Lx)  at (e1descsubcat.center |- p1.west);
\coordinate (e1Rx)  at (e1descrep.center    |- p1.west);
\path let \p{vec}=($(e1Rx)-(e1Lx)$),
          \n{g}={(\x{vec}-2.125cm)/3}
  in
  coordinate (e1Mx1) at ($(e1Lx)+(0.875cm+\n{g},0)$)
  coordinate (e1Mx2) at ($(e1Lx)+(1.50cm+2*\n{g},0)$);

\node[subcatSel] (e1s1) at ($(e1Lx)+(0,\elemTop)$) {Cat 1};
\node[subcat]    (e1s2) at ($(e1Lx)+(0,\elemCenter)$) {Cat 2};
\node[lab]                                 (e1s3) at ($(e1Lx)+(0,\elemBot)$) {$\vdots$};

\node[atk] (e1a1) at ($(e1Mx1)+(0,\elemTop+\pentaShift)$) {$A_1$};
\node[atk] (e1a2) at ($(e1Mx1)+(0,\elemCenter+\pentaShift)$) {$A_2$};
\node[atk] (e1a3) at ($(e1Mx1)+(0,\elemBot+\pentaShift)$) {$A_3$};
\node[caption,anchor=south] at ($(e1Mx1 |- p1.south)+(0,\captionOffset)$) {attack\\candidates};

\draw[zoom] (e1s1.north east) -- (e1a1.north);
\draw[zoom] (e1s1.south east) -- (e1a3.corner 3);

\node[llm] (e1llm) at ($(e1Mx2)+(0,\elemCenter)$) {LLM};

\node[atkDrop] (e1b1) at ($(e1Rx)+(0,\elemTop+\pentaShift)$) {$A_1$};
\node[atkSel]  (e1b2) at ($(e1Rx)+(0,\elemCenter+\pentaShift)$) {$A_2$};
\node[atkDrop] (e1b3) at ($(e1Rx)+(0,\elemBot+\pentaShift)$) {$A_3$};

\draw[arr] ($(e1a2.east)+(\arrowGap,-\pentaShift)$) -- ($(e1llm.west)+(-\arrowGap,0)$);
\draw[arr] ($(e1llm.east)+(\arrowGap,0)$)           -- ($(e1b2.west)+(-\arrowGap,-\pentaShift)$);
\node[arrowlab] at ($(e1Mx1)!0.5!(e1Rx)+(0,\elemTop)$) {(fairly compare ASR)};

\node[title] at ($(p2.north west)+(\titleInset,\titleDrop)$)
    {(P2) Initial selection of defenses};

\node[caption,anchor=south west] (e2descstages) at ($(p2.south west)+(\captionInset,\captionOffset)$)
    {defense\\pipeline stages};
\node[wideCap,anchor=south east] (e2descpchev) at ($(p2.south east)+(-\captionInset,\captionOffset)$)
    {per-stage\\ configs};

\coordinate (e2Lx)  at (e2descstages.center |- p2.west);
\coordinate (e2Rx)  at (e2descpchev.center  |- p2.west);
\path let \p{vec}=($(e2Rx)-(e2Lx)$),
          \n{g}={(\x{vec}-3cm)/4}
  in
  coordinate (e2Mx1) at ($(e2Lx)+(0.6cm+\n{g},0)$)
  coordinate (e2Mx2) at ($(e2Lx)+(0.85cm+2*\n{g},0)$)
  coordinate (e2Mx3) at ($(e2Lx)+(1.1cm+3*\n{g},0)$)
  coordinate (e2Mx4) at ($(e2Lx)+(1.35cm+4*\n{g},0)$);

\node[stageSel] (e2s1) at ($(e2Lx)+(0,\elemTop)$) {Stage 1};
\node[stage]    (e2s2) at ($(e2Lx)+(0,\elemCenter)$) {Stage 2};
\node[lab]                                (e2s3) at ($(e2Lx)+(0,\elemBot)$) {$\vdots$};

\node[def] (e2d1) at ($(e2Mx1)+(0,\elemTop)$) {$D_1$};
\node[def] (e2d2) at ($(e2Mx1)+(0,\elemCenter)$) {$D_2$};
\node[def] (e2d3) at ($(e2Mx1)+(0,\elemBot)$) {$D_3$};
\node[caption,anchor=south] at ($(e2Mx1 |- p2.south)+(0,\captionOffset)$) {defense\\candidates};

\draw[zoom] (e2s1.north east) -- (e2d1.north);
\draw[zoom] (e2s1.south east) -- (e2d3.south west);

\node[llm] (e2llm) at ($(e2Mx2)+(0,\elemCenter)$) {LLM};

\node[defSel]  (e2g1) at ($(e2Mx3)+(0,\elemTop)$) {$D_1$};
\node[defDrop] (e2r2) at ($(e2Mx3)+(0,\elemCenter)$) {$D_2$};
\node[defSel]  (e2g3) at ($(e2Mx3)+(0,\elemBot)$) {$D_3$};
\node[caption,anchor=south] at ($(e2Mx3 |- p2.south)+(0,\captionOffset)$) {high-utility\\candidates};

\node[llm] (e2llm2) at ($(e2Mx4)+(0,\elemCenter)$) {LLM};

\coordinate (e2c1) at ($(e2Rx)+(0,\elemTop)$);
\coordinate (e2c2) at ($(e2Rx)+(0,\elemCenter)$);
\coordinate (e2c3) at ($(e2Rx)+(0,\elemBot)$);

\chevAt[$D_1$]{e2c1}
\chevAt[$D_3$]{e2c2}
\chevAt[$\{D_1, D_3\}$]{e2c3}

\draw[arr] ($(e2d2.east)+(\arrowGap,0)$)  -- ($(e2llm.west)+(-\arrowGap,0)$);
\draw[arr] ($(e2llm.east)+(\arrowGap,0)$) -- ($(e2r2.west)+(-\arrowGap,0)$);
\node[arrowlab] at ($(e2Mx1)!0.5!(e2Mx3)+(0,\elemTop)$) {(compare utility)};

\draw[arr] ($(e2r2.east)+(\arrowGap,0)$)  -- ($(e2llm2.west)+(-\arrowGap,0)$);
\draw[arr] ($(e2llm2.east)+(\arrowGap,0)$) -- ($(e2c2)+({-\chevHalfW-\arrowGap},0)$);
\node[arrowlab] at ($(e2Mx4)!0.02!(e2c2)+(0,\elemTop)$) {(combine \& test)};

\node[title] at ($(p3.north west)+(\titleInset,\titleDrop)$)
    {(P3) Attacks versus defenses};

\node[lab,rotate=90] (e3s1) at ($(p3.west)+(0.45,0)$) {Stage 1};

\node[colCap,anchor=south west] (e3descacc) at ($(p3.south west)+(0.75,\captionOffset)$)
    {per-stage\\configs};
\node[wideCap,anchor=south east] (e3descbest) at ($(p3.south east)+(-\captionInset,\captionOffset)$)
    {preferred configs\\per-stage}; 

\coordinate (e3Lx) at (e3descacc.center  |- p3.west);
\coordinate (e3Rx) at (e3descbest.center |- p3.west);
\coordinate (e3Mx) at ($(e3Lx)!0.3!(e3Rx)$);

\coordinate (e3d1)  at ($(e3Lx)+(0,\elemTop)$);
\coordinate (e3d3)  at ($(e3Lx)+(0,\elemCenter)$);
\coordinate (e3d13) at ($(e3Lx)+(0,\elemBot)$);

\chevAtNoFill[$D_1$]{e3d1}
\chevAtNoFill[$D_3$]{e3d3}
\chevAtNoFill[$\{D_1, D_3\}$]{e3d13}

\node[boldlab] at ($($(e3Lx)+(\chevHalfW,0)$)!0.5!($(e3Mx)+(-0.24,0)$)+(0,\elemCenter)$) {VS};

\node[atkSel] (e3a1) at ($(e3Mx)+(0,\elemTop+\pentaShift)$) {$A_2$};
\node[atkSel] (e3a2) at ($(e3Mx)+(0,\elemCenter+\pentaShift)$) {$A_4$};
\node[lab] (e3an) at ($(e3Mx)+(0,\elemBot)$) {$\vdots$};
\node[caption,anchor=south] at ($(e3Mx |- p3.south)+(0,\captionOffset)$)
    {all attack\\representatives};

\coordinate (e3b1)   at ($(e3Rx)+(0,\elemTop)$);
\coordinate (e3b2)   at ($(e3Rx)+(0,\elemCenter)$);
\coordinate (e3best) at ($(e3Rx)+(0,\elemBot)$);

\chevDropAt[$D_1$]{e3b1}
\chevDropAt[$D_3$]{e3b2}
\chevAt[$\{D_1, D_3\}$]{e3best}

\node[llm] (e3llm) at ($($(e3Mx)+(0.24,0)$)!0.5!($(e3Rx)+(-\chevHalfW,0)$)+(0,\elemCenter)$) {LLM};
\draw[arr] ($(e3Mx)+(0.34,\elemCenter)$)        -- ($(e3llm.west)+(-\arrowGap,0)$);
\draw[arr] ($(e3llm.east)+(\arrowGap,0)$)       -- ($(e3Rx)+({-\chevHalfW-\arrowGap},\elemCenter)$);
\node[arrowlab] at ($(e3llm)+(0,\elemTop-\elemCenter)$) {(balance objectives)};

\node[title] at ($(p4.north west)+(\titleInset,\titleDrop)$)
    {(P4) Final cross-stage combination};

\node[caption,align=center,text width=1.6cm] (e4scenario) at ($(p4.west)+(1.05,\elemCenter+0.375)$)
    {Compare\\cross-stage\\defense\\combinations\\per scenario};
\draw[arr] ($(p4.west)+(2.0,\elemCenter)$) -- ($(p4.west)+(2.5,\elemCenter)$);

\node[colCap,anchor=south] (e4descsec) at ($(p4.south west)+(3.4,\captionOffset)$)
    {high\\security};
\node[colCap,anchor=south] (e4desceff) at ($(p4.south east)+(-2.0,\captionOffset)$)
    {high\\efficiency};

\coordinate (e4Lx) at (e4descsec.center |- p4.west);
\coordinate (e4Rx) at (e4desceff.center |- p4.west);
\coordinate (e4Mx) at ($(e4Lx)!0.5!(e4Rx)$);

\node[colCap,anchor=south] at ($(e4Mx |- p4.south)+(0,\captionOffset)$)
    {high\\utility};

\chevStackAt{e4sec}{e4Lx}
\chevAt[$\{D_1, D_3\}$]{e4sec1c}
\chevLLMAt{e4sec2c}
\chevAt[$D_5 \Rightarrow D_7$]{e4sec3c}

\chevStackAt{e4utl}{e4Mx}
\chevAt{e4utl1c}
\chevLLMAt{e4utl2c}
\chevAt[$D_8$]{e4utl3c}

\chevStackAt{e4eff}{e4Rx}
\chevAt[$\{D_1, D_3\}$]{e4eff1c}
\chevLLMAt{e4eff2c}
\chevAt{e4eff3c}

\coordinate (e4stagex) at ($($(e4Rx)+(\chevHalfW,0)$)!0.5!(p4.east)$);
\node[lab] (e4stage1lab) at (e4stagex |- e4eff1c) {Stage 1};
\node[lab] at (e4stagex |- e4eff2c) {$\vdots$};

\end{tikzpicture}
}
\caption{CASCADE's four sequential processes (P1--P4) select representative attacks (red pentagons), test them against per-stage defense configurations (blue chevrons) built from individual defenses (blue circles), and identify preferred cross-stage defense combinations for practical scenarios.\label{fig:framework}}
\end{figure}
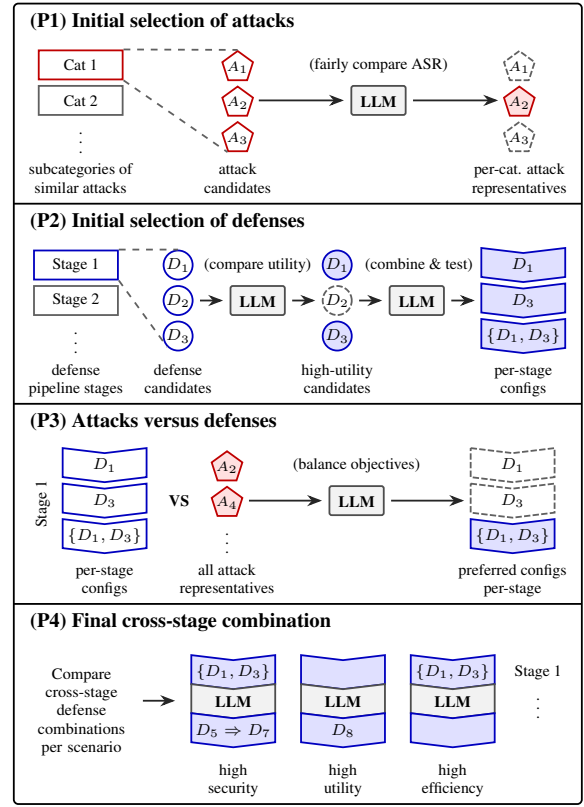 Jailbreak is considered an inherent LLM vulnerability, rooted in the tension between competing training objectives of \textit{helpfulness} and \textit{harmlessness}. More capable models can also follow malicious instructions more effectively, so safety must scale with capability \citep{wei2023jailbroken}. Despite safety alignment efforts (\citealp{touvron2023llama2openfoundation}; \citealp{openai2024gpt4technicalreport}), jailbreak risks persist as attackers continually evolve their strategies \citep{nasr2025attackermovessecondstronger}. These risks are \textit{no longer theoretical}: a user's suicide has prompted litigation over alleged safety failures in a deployed LLM system \citep{raine_openai_2025}. Substantial research has emerged on jailbreak attacks and defenses, yet three persistent gaps remain.

\noindent \textit{(1) Fair comparison} is hindered by widely varying experimental settings across studies \citep{chu-etal-2025-jailbreakradar}. Notably, the absence of unified metrics limits fair comparison across heterogeneous techniques \citep{chouldechova2025comparison}, a gap that persists in standardized evaluation benchmarks (\citealp{DBLP:conf/icml/MazeikaPYZ0MSLB24}; \citealp{DBLP:conf/nips/ChaoDRACSDFPTH024}) and empirical studies (\citealp{xu-etal-2024-comprehensive}; \citealp{shen2025pandaguardsystematicevaluationllm}).

\noindent \textit{(2) Practical defense} requires understanding which defenses work best under specific conditions and navigating trade-offs in security, efficiency, and utility (\citealp{shen2025pandaguardsystematicevaluationllm}; \citealp{11573588}). Despite broad coverage, these studies offer high-level insights rather than concrete deployment recommendations, and none examine how defenses should be combined.

\noindent \textit{(3) Defense combinations}, both within and across pipeline stages, have not been systematically studied. Though no single defense is universally effective (\citealp{shen2025pandaguardsystematicevaluationllm}; \citealp{chu-etal-2025-jailbreakradar}), defenses are evaluated only in isolation, leaving effective combinations a key open problem.

Hence, we propose CASCADE, to our knowledge, the first framework for fair, controlled, systematic evaluation of defense combinations across pipeline stages, toward practical defense recommendations. Our contributions are as follows:
\begin{itemize}
\item \textbf{Fair comparison framework.} We propose a decision framework with well-defined metrics, standardized experiment settings, and explicit fairness rules for the controlled evaluation of jailbreak attacks and defenses.
    \item \textbf{Practical defense combinations.} We systematically evaluate defense combinations within and across pipeline stages, identifying those that achieve substantial security with minimal utility or efficiency degradation.
    \item \textbf{Empirical evaluation.} We evaluate \numattacks unique attacks and \numdefenses unique defenses under our framework, providing actionable recommendations for practical adoption and open-sourcing our code\footnote{\url{https://github.com/Singa-pirate/CASCADE-jailbreak-defense-combi}} to support further research.
\end{itemize}

\section{Background and Related Work}

Contemporary jailbreaks span diverse attack surfaces, including multi-turn conversations \citep{307992} and LLM-based agents \citep{DBLP:conf/iclr/AndriushchenkoS25}. We focus on the foundational primitive of \textit{direct}, \textit{black-box}, \textit{single-turn} jailbreaks, where a single adversarial prompt attempts to bypass the safety measures of a black-box LLM. Techniques applied in this restrictive setting remain applicable in less restrictive ones, making it the most actionable regime for practical defense recommendations. 

In \textit{direct} attacks, the user is the attacker, submitting the malicious prompt themselves rather than via injected external content \citep{vassilev2025adversarial}. In the \textit{black-box} setting, the attacker has only query access to the target LLM, observing its text output without visibility into model internals such as weights, gradients, or logits \citep{yi2024jailbreakattacksdefenseslarge}. \textit{Single-turn} attacks succeed in a single query, without exploiting conversation history or multi-turn escalation \citep{307992}. 

\subsection{Jailbreak Attacks}

\providecommand{\colSpacing}{2.25}
\providecommand{\leafGap}{-0.44}

\begin{figure}
    \centering
    \resizebox{\columnwidth}{!}{
    \begin{tikzpicture}[
        font=\rmfamily\footnotesize,
        category/.style={
            thick,
            rounded corners=2pt,
            draw=\neutralColor,
            align=center,
            font=\rmfamily\scriptsize,
            minimum width=2cm,
            minimum height=0.68cm,
            inner sep=2pt
        },
        leaf/.style={
            thick,
            rounded corners=2pt,
            draw=\attackColor,
            align=left,
            font=\rmfamily\tiny,
            inner sep=2.5pt,
            outer sep=0pt,
            minimum width=2cm,
            minimum height=1.3cm,
            text width=1.82cm
        },
        connector/.style={
            thick,
            draw=\arrowColor,
            rounded corners=2pt
        }
    ]

\def\colA{0}
    \pgfmathsetmacro{\colB}{\colSpacing}
    \pgfmathsetmacro{\colC}{2*\colSpacing}

\node[category] (white)      at (\colA,0) {White-box\\transferable};
    \node[category] (black_temp) at (\colB,0) {Black-box\\template-based};
    \node[category] (black_llm)  at (\colC,0) {Black-box\\LLM-based};

\node[leaf, anchor=north] (list_white) at (\colA,\leafGap) {
        \textbullet~[GCG]\\
        \textbullet~[AmpleGCG]\\
        \textbullet~[I-GCG]\\
        \textbullet~[DSN]\\
        \textbullet~[Adaptive]
    };

    \node[leaf, anchor=north] (list_temp) at (\colB,\leafGap) {
        \textbullet~[Prefix]\\
        \textbullet~[Refusal]\\
        \textbullet~[Wiki]\\
        \textbullet~[DAN]\\
        \textbullet~[DevMode]\\
        \textbullet~[AIM]\\
        \textbullet~[Code]\\
        \textbullet~[Flip]\\
        \textbullet~[MultiJail]\\
        \textbullet~[SeqBreak]
    };

\node[leaf, anchor=north, minimum height=0pt] (list_llm) at (\colC,\leafGap) {
        \textbullet~[TAP]\\
        \textbullet~[PAP]\\
        \textbullet~[ReNeLLM]\\
        \textbullet~[GPTFuzzer]
    };

\draw[connector] (white.south)      -- (list_white.north);
    \draw[connector] (black_temp.south) -- (list_temp.north);
    \draw[connector] (black_llm.south)  -- (list_llm.north);

    \end{tikzpicture}
    }
    \captionsetup{
        justification=justified,
    }
    \caption{Three jailbreak attack categories (in-scope) and \numattacks unique techniques evaluated; details in Table~\ref{tab:appendix attacks}.}
    \label{fig:attacks}
\end{figure}
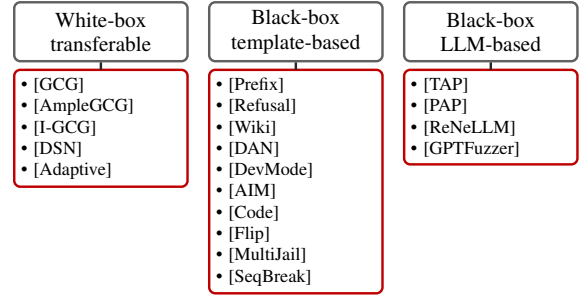
 
Attacks commonly divide into white-box and black-box, the latter including template-based and LLM-based approaches \citep{yi2024jailbreakattacksdefenseslarge}. White-box techniques require model internals beyond our threat model, but we include the \textit{transferable} prompts they produce, which are applicable to the black-box setting. Figure~\ref{fig:attacks} summarizes the scope, with each category detailed below.

\noindent \textit{(1) White-box transferable} attacks involve a surrogate LLM's gradients to optimize for jailbreak success, generating prompts to attack the target LLM. For example, GCG refines a universal suffix to elicit affirmative responses \citep{zou2023universaltransferableadversarialattacks}.

\noindent \textit{(2) Black-box template-based} attacks use templates: output style constraints \citep{wei2023jailbroken}, fictional personas \citep{10.1145/3658644.3670388}, or structured formats (\citealp{liu2025flipattack}; \citealp{saiem-etal-2025-sequentialbreak}). For example, CodeAttack disguises the malicious prompt as a code-completion task \citep{ren-etal-2024-codeattack}.

\noindent \textit{(3) Black-box LLM-based} attacks automate prompt generation via red-team LLMs (i.e., LLMs acting as adversaries). For example, TAP uses attacker and evaluator LLMs to refine prompts in a branch-evaluate-prune cycle \citep{DBLP:conf/nips/MehrotraZKNASK24}.

\subsection{Jailbreak Defenses}

Adapting the stage-based taxonomies of prior work (\citealp{chen2025teleaisafetycomprehensivellmjailbreaking}; \citealp{11573588}), we organize defenses (Figure~\ref{fig:defenses}), focusing on the three stages easiest to apply in practice; stages detailed below.

\noindent \textit{(1) Training} includes data cleaning and post-training safety alignment \citep{wang2025comprehensivesurveyllmagentstack}.

\noindent \textit{(2) Input guard} screens and blocks harmful inputs, using signals such as perplexity \citep{jain2023baselinedefensesadversarialattacks} or guard models such as WildGuard \citep{DBLP:conf/nips/HanREJL00D24} and PromptGuard \citep{huggingfaceMetallamaLlamaPromptGuard286MHugging}.

\noindent \textit{(3) Input modification} transforms the user prompt to neutralize harmful intent. Examples range from safety system prompts \citep{Xie2023Defending} to optimized defense strings such as RPO \citep{zhou2024robust} and DPP \citep{xiong-etal-2025-defensive}.

\noindent \textit{(4) Inference} intervenes within the model at inference time, for example by inspecting safety-critical parameter gradients or steering decoding (\citealp{xie-etal-2024-gradsafe}; \citealp{xu-etal-2024-safedecoding}).

\noindent \textit{(5) Output refinement} post-processes the output, often using secondary LLMs to inspect, revise, or reject the final response \citep{wang-etal-2024-defending}.

\noindent \textit{(6) Output guard} screens query-response pairs and blocks harmful conversations, using guard models such as LlamaGuard \citep{inan2023llamaguardllmbasedinputoutput} or reasoning models such as GuardReasoner \citep{liu2025guardreasonerreasoningbasedllmsafeguards}. Some input guards such as WildGuard also apply here.

\providecommand{\stagePitch}{1.1}
\providecommand{\horizGap}{0.3}
\providecommand{\labelShift}{0.2}
\providecommand{\pipelineX}{-1.0}
\providecommand{\labelHeight}{0.3cm}

\begin{figure}
    \centering
    \resizebox{\columnwidth}{!}{
    \begin{tikzpicture}[
        font=\rmfamily\footnotesize,
        stage/.style={
            thick,
            rounded corners=2pt,
            draw=\neutralColor,
            align=center,
            font=\rmfamily\scriptsize,
            minimum width=2.2cm,
            minimum height=0.8cm,
            inner sep=2pt
        },
        pipelabel/.style={
            font=\rmfamily\scriptsize,
            minimum height=\labelHeight
        },
        tech_list_selected/.style={
            thick,
            rounded corners=2pt,
            draw=\defenseColor,
            align=left,
            font=\rmfamily\tiny,
            minimum width=1.0cm,
            inner sep=2pt,
            outer sep=0pt,
            text width=1.0cm
        },
        flow_arrow/.style={
            -{Stealth[length=1.5mm,width=1.2mm]},
            thick,
            draw=\arrowColor
        },
        link_line/.style={
            thick,
            draw=\arrowColor
        }
    ]

\node[pipelabel] (user_in) at (\pipelineX,  0) {User prompt};
    \node[stage] (if)  at (\pipelineX, {-1*\stagePitch+\labelShift}) {(2)\\Input guard};
    \node[stage] (im)  at (\pipelineX, {-2*\stagePitch+\labelShift}) {(3)\\Input modification};
    \node[stage] (inf) at (\pipelineX, {-3*\stagePitch+\labelShift}) {(4)\\Inference};
    \node[stage] (or)  at (\pipelineX, {-4*\stagePitch+\labelShift}) {(5)\\Output refinement};
    \node[stage] (of)  at (\pipelineX, {-5*\stagePitch+\labelShift}) {(6)\\Output guard};
    \node[pipelabel] (sys_out) at (\pipelineX, {-6*\stagePitch+2*\labelShift}) {Final response};

\draw[flow_arrow] (user_in) -- (if);
    \draw[flow_arrow] (if) -- (im);
    \draw[flow_arrow] (im) -- (inf);
    \draw[flow_arrow] (inf) -- (or);
    \draw[flow_arrow] (or) -- (of);
    \draw[flow_arrow] (of) -- (sys_out);

\node[stage, anchor=east] (tr) at ($(inf.west)+(-\horizGap,0)$) {(1)\\Training};

    \draw[flow_arrow] (tr) -- (inf);

\node[tech_list_selected, anchor=west] (list_if) at ($(if.east)+(\horizGap,0)$) {
        \textbullet~[PPL]\\
        \textbullet~[WG]\\
        \textbullet~[W-PPL]\\
        \textbullet~[OSS]\\
        \textbullet~[PG]\\
        \textbullet~[Mod]
    };

    \node[tech_list_selected, anchor=west] (list_of) at ($(of.east)+(\horizGap,0)$) {
        \textbullet~[LG]\\
        \textbullet~[XG]\\
        \textbullet~[WG]\\
        \textbullet~[GR]\\
        \textbullet~[Mod]
    };

    \node[tech_list_selected] (list_im) at ($(list_if.south)!0.5!(list_of.north)$) {
        \textbullet~[S-LLM]\\
        \textbullet~[PAT]\\
        \textbullet~[SR]\\
        \textbullet~[DPP]\\
        \textbullet~[ICD]\\
        \textbullet~[RPO]
    };

\draw[link_line] (if.east) -- (list_if.west);
    \draw[link_line] (im.east) -- (list_im.north);
    \draw[link_line] (of.east) -- (list_of.west);

    \end{tikzpicture}
    }
    \captionsetup{
        justification=justified,
    }
    \caption{The six-stage LLM defense pipeline, with \numdefenses unique contemporary defense techniques shown for the three in-scope stages (input guard, input modification, output guard); WG and Mod serve dual roles in stages (2) and (6); details in Table~\ref{tab:appendix defenses}.}
    \label{fig:defenses}
\end{figure}
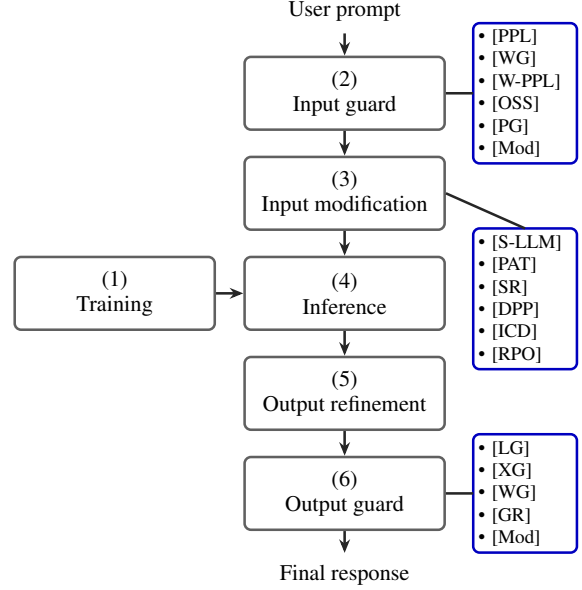
 
\subsection{Evaluation Methods and Metrics}
\label{subsec:Evaluation methods and metrics}

\paragraph{Jailbreak effectiveness} is commonly evaluated on a malicious-goal dataset, with a judge scoring jailbreak attempts. Generally, \textit{Attack Success Rate (ASR)} describes the proportion of jailbroken goals:
\begin{equation}
\small
  \label{eq:asr}
  ASR = \frac{\sum_{d} I(Q_d)}{ |D|},
\end{equation}
where $Q_d$ is the attempt for the $d$-th malicious goal in dataset $D$, and the indicator $I$, scored by a judge, returns $1$ for a successful jailbreak and $0$ otherwise. Higher ASR indicates a more effective attack or a less effective defense, and vice versa. ASR is widely used, but varying instantiations could cause unfair comparisons \citep{chouldechova2025comparison}:

\noindent \textit{(1) ASR\textsubscript{avg}} averages success over $N$ attempts per malicious goal \citep{zou2023universaltransferableadversarialattacks}, where $n \in \{1,\ldots,N\}$ indexes attempts:
\begin{equation}
\small
  \label{eq:asr_average}
  ASR_{avg} = \frac{\sum_{n}\sum_{d} I(Q_{d,n})}{ |D| * N}.
\end{equation}

\noindent \textit{(2) ASR\textsubscript{@N}}, or top-N ASR, considers a goal to be jailbroken if any of the $N$ attack attempts succeeds \citep{zhou-etal-2025-dont}:
\begin{equation}
\small
  \label{eq:asr_max}
  ASR_{@N} = \frac{\sum_{d} \max_{n}[I(Q_{d,n})]}{|D|}.
\end{equation}

\noindent \textit{(3) ASR\textsubscript{ensemble}} treats a goal as jailbroken if any of $V$ attack variants succeeds \citep{ding-etal-2024-wolf}, where $v \in \{1,\ldots,V\}$ indexes variants:
\begin{equation}
\small
  \label{eq:asr_ensemble}
  ASR_{ens} = \frac{\sum_{d} \max_{v}[I(Q_{d,v})]}{ |D|}.
\end{equation}

Beyond instantiation choice, ASR accuracy depends on the judge that scores each attempt. Early rule-based judges, such as keyword filters, produce frequent misjudgments. Recent LLM-based judges include GPT-4 evaluator \citep{DBLP:conf/iclr/Qi0XC0M024}, HarmBench judge \citep{DBLP:conf/icml/MazeikaPYZ0MSLB24}, ShieldLM \citep{zhang-etal-2024-shieldlm}, and fine-tuned RoBERTa models \citep{xu-etal-2024-comprehensive}, but still produce false positives and negatives (\citealp{shen2025pandaguardsystematicevaluationllm}; \citealp{chouldechova2025comparison}). Despite proposals such as majority voting over an ensemble of judges \citep{zhou-etal-2025-dont}, no strategy clearly aligns best with human judgment.

ASR also depends on the dataset $D$ of malicious goals, which spans major AI safety policy categories. AdvBench is an early benchmark \citep{zou2023universaltransferableadversarialattacks}, though later work notes duplicated entries \citep{DBLP:conf/nips/ChaoDRACSDFPTH024}. Recent alternatives include HarmBench \citep{DBLP:conf/icml/MazeikaPYZ0MSLB24} and JBB-Behaviors \citep{DBLP:conf/nips/ChaoDRACSDFPTH024}.

\paragraph{LLM response quality} is also important, as defenses can degrade responses due to over-refusal. \textit{Utility} metrics measure a defended LLM's ability to answer normal queries. A common utility framework, AlpacaEval \citep{dubois2025lengthcontrolledalpacaevalsimpleway}, reports the target LLM's length-controlled win-rate against a reference model under LLM-based judging. Other datasets, including XSTest \citep{rottger-etal-2024-xstest} and OR-Bench \citep{DBLP:conf/icml/CuiCSH25}, further benchmark over-refusal on benign prompts near the safety boundary.

\subsection{Empirical Studies}

Several empirical studies have evaluated jailbreak attacks and defenses. \citet{xu-etal-2024-comprehensive} provide an early evaluation of 9 attacks and 7 defenses. PandaGuard evaluates 19 attacks against 9 defenses (out of 12 implemented), finding no universally best defense \citep{shen2025pandaguardsystematicevaluationllm}. The Security-Efficiency-Utility framework reaches the same conclusion across 9 attacks and 9 guardrail defenses \citep{11573588}. AISafetyLab \citep{zhang2025aisafetylabcomprehensiveframeworkai} and TeleAI-Safety \citep{chen2025teleaisafetycomprehensivellmjailbreaking} further expand coverage. While these studies consistently find no universally optimal defense, systematic evaluation of defense combinations remains unexplored. Our work addresses the gap with the first systematic study, to our knowledge, of defense combinations within and across pipeline stages, yielding practical recommendations for layered defense pipelines.

\section{Methodology}

Our CASCADE framework consists of four sequential processes that progressively refine the technique set, illustrated in Figure~\ref{fig:framework}. The rules below define one fair selection procedure that practitioners can adapt for different use cases; rationales for each rule are given in Appendix~\ref{sec:appendix:b}.

\noindent \textit{(P1) Initial selection of attacks.} We group attacks into \textit{subcategories} based on similarity. Within each subcategory, we evaluate ASR on target LLMs and select effective attacks to represent the group. This grouping improves experimental efficiency while ensuring coverage of diverse attack patterns. To ensure fairness, we assume the \textbf{same attacker capabilities} for all attacks (per the threat model) and apply the \textbf{same target LLM query budget per subcategory} (reflecting practical query cost and detection risk).

\noindent \textit{(P2) Initial selection of defenses.} We group defenses by pipeline stage. Within each \textit{stage}, we evaluate the utility of individual defenses and intra-stage defense combinations, eliminating those that cause unacceptable utility degradation, via \textbf{top-k rank utility selection} within reasonable groupings of defense configurations.

\noindent \textit{(P3) Attacks versus defenses.} For each pipeline stage, we test selected defenses against the attack representatives to evaluate ASR reduction, selecting preferred defense configurations within the search space. We give attacks the \textbf{same query budgets as before} (simulating effective attack scenarios) and select based on \textbf{balanced performance} across ASR reduction, utility preservation, and memory use.

\noindent \textit{(P4) Final cross-stage combination.} We evaluate cross-stage combinations of the preferred defenses using the same \textbf{top-k rank utility selection} and \textbf{ASR evaluation}, identifying recommended defense combinations for practical scenarios.

\subsection{Metrics}

To address the unfair comparisons caused by varying ASR definitions (Section~\ref{subsec:Evaluation methods and metrics}; \citealp{chouldechova2025comparison}), we adopt a unified formulation applicable across techniques. 

\paragraph{Jailbreak effectiveness} is measured by $ASR_{@max\_q}$, combining $ASR_{@N}$ and $ASR_{ens}$ for fair application across techniques. We fix a maximum query budget per subcategory, so attacks with more variants receive fewer repetitions:
\begin{equation}
\small
ASR_{@max\_q} = \frac{\sum_{d}\max_{n, v}[I(Q_{d,n,v})]}{|D|},
\end{equation}
where $Q_{d,n,v}$ denotes the $n$-th attempt of the $v$-th variant for the $d$-th malicious goal. $ASR_{@max\_q}$ captures jailbreak success within the budget.

\paragraph{LLM response quality} is measured by AlpacaEval's length-controlled win-rate:
\begin{equation}
\small
\begin{aligned}
Utility &:= \mathrm{WinRate}^{\mathrm{LC}}(m,b) \\
               &= 100 \cdot \mathbb{E}_x \left[
\sigma\big(
Q_{m,b}
+ \underbrace{0}_{\text{length term}}
+ D_{m,b,x}
\big)
\right]
\end{aligned},
\end{equation}
where $m$ is the target model, $b$ the baseline reference, $x$ a normal instruction, and $\sigma$ a logistic regression over model quality $Q$, length (set to zero), and instruction difficulty $D$.

\paragraph{Defense effectiveness} is measured by two aggregates that we seek to jointly maximize, subject to an \emph{inherent tradeoff} between them. $\resultASR$ averages the percentage reduction of $ASR_{@max\_q}$ across attack representatives:
\begin{equation}
\small
\resultASR = \frac{1}{|A|}\sum_a\frac{ASR_a-ASR_{a}'}{ASR_a}
\end{equation}
where $a$ indexes representative attacks in $A$, and $ASR_a$, $ASR_{a}'$ are its $ASR_{@max\_q}$ before and after the defense. Symmetrically, $\resultU$ is the average percentage increase of utility across target LLMs:
\begin{equation}
\small
\resultU = \frac{1}{|M|}\sum_m\frac{U_m'-U_{m}}{U_m}
\end{equation}
where $m$ indexes target LLMs in $M$, and $U_m$, $U_m'$ are its utility before and after the defense.

\subsection{Implementation and Experimental Setup}

Our pipeline-based architecture (Figure~\ref{fig:system-design}) constructs an execution graph of attack, defense, and target LLM nodes; each node updates a shared state object that records the prompt, response, rejection status, and evaluation result. We summarize three key settings below, with full details in Appendix~\ref{subsec:appendix:a.3}.

\noindent \textit{(1) Datasets and judges.} For jailbreak evaluation, we use the JBB-Behaviors dataset (100 malicious goals; adopted by \citealp{zhou2024robust}; \citealp{robey2024smoothllmdefendinglargelanguage}) paired with the HarmBench judge (as in \citealp{zhou-etal-2025-dont}), balancing quality, efficiency, and prior adoption. For utility, we use the AlpacaEval framework with Llama-3.1-70B-Instruct as the LLM judge, following a recommended AlpacaEval setting for strong human agreement and efficiency.

\noindent \textit{(2) Target LLMs.} We select target LLMs per experiment, covering nine open-weight and proprietary models in total (Table~\ref{tab:appendix LLMs}). We disable reasoning for target LLMs to isolate base safety behavior under identical generation conditions.

\noindent \textit{(3) LLM settings.} We set the temperature to 0 with a constant seed for all judges, and for LLM generations in utility evaluations; 1 for target LLMs in repeated jailbreak attempts. We set max\_new\_tokens to 2048 (non-reasoning models) and 4096 (reasoning models) to avoid unexpected output truncations.

\providecommand{\sysMidX}{0}

\providecommand{\sysChevHalfW}{1.02}

\providecommand{\sysArrLen}{0.3}

\providecommand{\sysGoalY}{0.00}

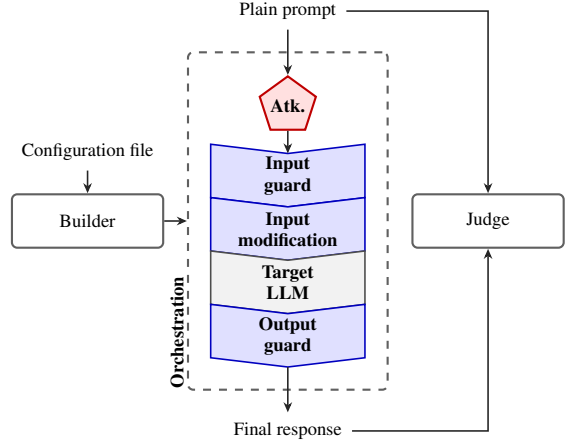
\begin{figure}[t]
    \centering
    \begin{tikzpicture}[
        font=\rmfamily\scriptsize,
artifact/.style={
            thick,
            rounded corners=2pt,
            draw=\neutralColor,
            align=center,
            font=\rmfamily\scriptsize,
            minimum width=2cm,
            minimum height=0.706cm,
            inner sep=2pt
        },
plain/.style={
            font=\rmfamily\scriptsize,
            minimum width=0.825cm,
            minimum height=0.3cm,
            align=center
        },
        atkSel/.style={
            pentagonBase,
            draw=\attackColor,
            fill=\attackFill,
minimum size=7.81mm,
            font=\rmfamily\scriptsize\bfseries
        },
        chevDef/.style={
            semithick,
            line join=miter,
            draw=\defenseColor,
            fill=\defenseFill
        },
        chevLLM/.style={
            semithick,
            line join=miter,
            draw=\llmColor,
            fill=\llmFill
        },
        arr/.style={
            -{Stealth[length=1.35mm,width=1.15mm]},
            semithick,
            draw=\arrowColor
        }
    ]

\pgfmathsetmacro{\chevPitch}{\sysChevHalfW*(\chevAspectH+\chevAspectN)}
    \pgfmathsetmacro{\sysOuterArrLen}{2*\sysArrLen}

\node[plain] (goal) at (\sysMidX, \sysGoalY) {Plain prompt};

\node[atkSel, anchor=north] (attacks) at ($(goal.south)+(0,-\sysOuterArrLen)$) {Atk.};

\coordinate (inputGuardC) at ($(attacks.south)+(0,{-\sysArrLen-\sysChevHalfW*\chevAspectN})$);
    \coordinate (inputModC)    at ($(inputGuardC)+(0,-\chevPitch)$);
    \coordinate (llmC)         at ($(inputModC)+(0,-\chevPitch)$);
    \coordinate (outputGuardC) at ($(llmC)+(0,-\chevPitch)$);

\node[plain, anchor=north] (response) at
        ($(outputGuardC)+(0,{-\sysChevHalfW*\chevAspectH-\sysOuterArrLen})$) {Final response};

    \drawChevron{chevDef}{inputGuardC}{\sysChevHalfW}
    \node[boldlab] at (inputGuardC) {Input\\guard};

    \drawChevron{chevDef}{inputModC}{\sysChevHalfW}
    \node[boldlab] at (inputModC) {Input\\modification};

    \drawChevron{chevLLM}{llmC}{\sysChevHalfW}
    \node[boldlab] at (llmC) {Target\\LLM};

    \drawChevron{chevDef}{outputGuardC}{\sysChevHalfW}
    \node[boldlab] at (outputGuardC) {Output\\guard};

\coordinate (inputGuardTopLeft)  at ($(inputGuardC)+(-\sysChevHalfW,{\sysChevHalfW*\chevAspectH})$);
    \coordinate (inputGuardTopRight) at ($(inputGuardC)+(\sysChevHalfW,{\sysChevHalfW*\chevAspectH})$);
    \coordinate (outputGuardBotLeft)  at ($(outputGuardC)+(-\sysChevHalfW,{-\sysChevHalfW*\chevAspectN})$);
    \coordinate (outputGuardBotRight) at ($(outputGuardC)+(\sysChevHalfW,{-\sysChevHalfW*\chevAspectN})$);
\coordinate (outputGuardBotPoint) at ($(outputGuardC)+(0,{-\sysChevHalfW*\chevAspectH})$);

\draw[arr] (goal.south)        -- (attacks.north);
    \draw[arr] (attacks.south)     -- ($(inputGuardC)+(0,{\sysChevHalfW*\chevAspectN})$);
    \draw[arr] ($(outputGuardC)+(0,{-\sysChevHalfW*\chevAspectH})$) -- (response.north);

\begin{scope}[on background layer]
        \node[
            draw=\neutralColor,
            fill=none,
            dashed,
            thick,
            rounded corners=2pt,
            inner sep=\sysArrLen cm,
            fit=(attacks) (inputGuardTopLeft) (inputGuardTopRight)
                (outputGuardBotLeft) (outputGuardBotRight) (outputGuardBotPoint)
        ] (orch_box) {};
    \end{scope}

    \node[
        boldlab,
        rotate=90,
        anchor=south west,
        inner sep=1pt
    ] at (orch_box.south west) {Orchestration};

\coordinate (sideMid) at ($(goal)!0.5!(response)$);

\path let \p1 = (orch_box.west), \p2 = (sideMid) in
        node[artifact, anchor=east] (builder) at (\x1-\sysArrLen cm, \y2) {Builder};
    \node[plain, anchor=south] (config) at ($(builder.north)+(0,\sysArrLen)$) {Configuration file};
    \draw[arr] (config)       -- (builder);
    \draw[arr] (builder.east) -- (orch_box.west |- builder.east);

\path let \p1 = (orch_box.east), \p2 = (sideMid) in
        node[artifact, anchor=west] (eval) at (\x1+\sysArrLen cm, \y2) {Judge};
    \draw[arr] (goal.east)     -| (eval.north);
    \draw[arr] (response.east) -| (eval.south);

    \end{tikzpicture}

    \captionsetup{
        justification=justified,
    }
    \caption{Pipeline-based system architecture: the Builder (left) wires the Orchestration framework (center) from a configuration file, and the Judge (right) scores the final response.}
    \label{fig:system-design}
\end{figure}

\section{Results}

\begin{table*}[t]
\centering
\footnotesize
\renewcommand{\arraystretch}{1.1}
\setlength{\tabcolsep}{2pt}
\newcommand{\headerpad}{\rule{0pt}{3.5ex}}
\begin{tabular*}{\textwidth}{@{\extracolsep{\fill}}lccccccccc}
\toprule
\multirow{2}{*}{\headerpad\textbf{Subcategory}} &
\multirow{2}{*}{\rule{0pt}{4.2ex}\makecell[c]{\textbf{Query}\\ \textbf{Budget}}} &
\multirow{2}{*}{\headerpad\textbf{Attack}} &
\multicolumn{6}{c}{\textbf{Target LLM}} &
\multirow{2}{*}{\headerpad\textbf{Average}} \\
\cmidrule(lr){4-9}
& & & \textbf{Vicuna} & \textbf{Llama2} & \textbf{Llama3} & \textbf{GPT-3.5} & \textbf{GPT-4o} & \textbf{Claude-sonnet-4} & \\
\midrule

\multirow{6}{*}{\makecell[l]{White-box\\transferable}}
& \multirow{6}{*}{15}
& \textbf{Adaptive} & \textbf{100.00} & 2.00 & \textbf{100.00} & \textbf{100.00} & 9.00 & \textbf{44.00} & \textbf{59.17} \\
& & \textbf{DSN} & 93.00 & \textbf{97.00} & 30.00 & 94.00 & 13.00 & 2.00 & 54.83 \\
& & GCG & \textbf{100.00} & 42.00 & 40.00 & 99.00 & 15.00 & 2.00 & 49.67 \\
& & AmpleGCG & 98.00 & 19.00 & 26.00 & 98.00 & \textbf{16.00} & 5.00 & 43.67 \\
& & I-GCG & 91.00 & 3.00 & 8.00 & 74.00 & 8.00 & 2.00 & 31.00 \\
& & Baseline & 83.00 & 7.00 & 22.00 & 56.00 & 7.00 & 5.00 & 30.00 \\

\midrule
\multirow{4}{*}{\makecell[l]{Black-box\\template-based,\\output style pattern}}
& \multirow{4}{*}{5}
& \textbf{Refusal} & 94.00 & \textbf{19.00} & \textbf{46.00} & \textbf{92.00} & \textbf{38.00} & \textbf{8.00} & \textbf{49.50} \\
& & Prefix & \textbf{99.00} & 7.00 & 10.00 & 87.00 & 6.00 & 4.00 & 35.50 \\
& & Wiki & 80.00 & 9.00 & 7.00 & 62.00 & 5.00 & 0.00 & 27.17 \\
& & Baseline & 58.00 & 4.00 & 16.00 & 46.00 & 7.00 & 5.00 & 22.67 \\

\midrule
\multirow{4}{*}{\makecell[l]{Black-box\\template-based,\\persona pattern}}
& \multirow{4}{*}{5}
& DevMode & 99.00 & \textbf{9.00} & \textbf{59.00} & \textbf{64.00} & 0.00 & 0.00 & \textbf{38.50} \\
& & DAN & 95.00 & 8.00 & 57.00 & 33.00 & 0.00 & 1.00 & 32.33 \\
& & AIM & \textbf{100.00} & 5.00 & 33.00 & 2.00 & 0.00 & 0.00 & 23.33 \\
& & Baseline & 58.00 & 4.00 & 16.00 & 46.00 & \textbf{7.00} & \textbf{5.00} & 22.67 \\

\midrule
\multirow{5}{*}{\makecell[l]{Black-box\\template-based,\\disguise pattern}}
& \multirow{5}{*}{36}
& \textbf{SeqBreak} & \textbf{100.00} & 97.00 & \textbf{100.00} & \textbf{100.00} & \textbf{98.00} & \textbf{60.00} & \textbf{92.50} \\
& & Code & 99.00 & \textbf{100.00} & \textbf{100.00} & \textbf{100.00} & 97.00 & 24.00 & 86.67 \\
& & Flip & 98.00 & 58.00 & 81.00 & \textbf{100.00} & 80.00 & 0.00 & 69.50 \\
& & MultiJail & \textbf{100.00} & 71.00 & 88.00 & 95.00 & 0.00 & 0.00 & 59.00 \\
& & Baseline & 94.00 & 9.00 & 26.00 & 63.00 & 8.00 & 6.00 & 34.33 \\

\midrule
\multirow{5}{*}{\makecell[l]{Black-box\\LLM-based}}
& \multirow{5}{*}{80}
& \textbf{ReNeLLM} & 91.00 & \textbf{81.00} & 91.00 & \textbf{95.00} & \textbf{87.00} & 12.00 & \textbf{76.17} \\
& & PAP & 70.00 & 42.00 & \textbf{94.00} & 59.00 & 41.00 & 3.00 & 51.50 \\
& & GPTFuzzer & 73.00 & 39.00 & 79.00 & 69.00 & 34.00 & 1.00 & 49.17 \\
& & TAP & 75.00 & 34.00 & 33.00 & 78.00 & 36.00 & \textbf{17.00} & 45.50 \\
& & Baseline & \textbf{98.00} & 9.00 & 27.00 & 67.00 & 11.00 & 9.00 & 36.83 \\

\bottomrule
\end{tabular*}
\captionsetup{
    justification=justified,
}
\caption{(P1) $ASR_{@max\_q}$ results (\%) across attacks and target LLMs. (Baseline: plain malicious strings; bold values: subcategory maximum; bold attack names: selected.)}
\label{tab:initial_attack_selection}
\end{table*} 

\subsection{Initial Attack Selection Results}

In the first process, we evaluate attacks on six LLMs commonly used in prior work: three open-weight (Vicuna, Llama2, Llama3) and three proprietary models (GPT-3.5, GPT-4o, Claude-sonnet-4). For each subcategory, we set a query budget divisible by each attack's variant count, while ensuring sufficient repetitions per variant. From the results (Table~\ref{tab:initial_attack_selection}), we select a minimal set of effective attacks per subcategory. Among white-box transferable attacks, we select both \textbf{Adaptive} and \textbf{DSN} for their strong overall performance. Persona-pattern attacks under-perform the baseline on advanced LLMs (GPT-4o and Claude-sonnet-4), likely due to modern safety training against fixed personas; we therefore exclude this subcategory. This yields five attack representatives: \textbf{Adaptive}, \textbf{DSN}, \textbf{Refusal}, \textbf{SeqBreak}, and \textbf{ReNeLLM}.

\subsection{Initial Defense Selection Results}
\label{subsec:initial defense}

\begin{figure}[t]
    \centering
    \includegraphics[width=0.48\textwidth]{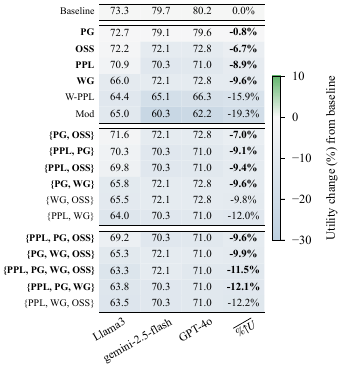}
    \captionsetup{
        justification=justified,
}
    \captionof{figure}{(P2) Utility across \emph{input guard} configurations. (Baseline: no defense; bold: selected.)}
    \label{fig:utility-heatmap-input-guard}
\end{figure}

\begin{figure}[t]
    \centering
    \includegraphics[width=0.462\textwidth]{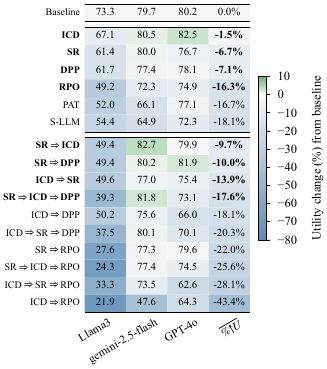}
    \captionsetup{
        justification=justified,
}
    \captionof{figure}{(P2) Utility across \emph{input modification} configurations. (Baseline: no defense; bold: selected.) DPP and RPO both add an optimized defense suffix, so we exclude combinations containing both.}
    \label{fig:utility-heatmap-input-modification}
\end{figure}

\begin{figure}[t]
    \centering
    \includegraphics[width=0.48\textwidth]{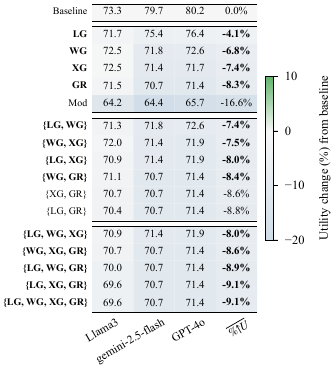}
    \captionsetup{
        justification=justified,
}
    \captionof{figure}{(P2) Utility across \emph{output guard} configurations. (Baseline: no defense; bold: selected.)}
    \label{fig:utility-heatmap-output-guard}
\end{figure}

In the second process, we evaluate the utility of defended LLMs, using newer LLMs popular in practice at the time of writing (informed by sources such as OpenRouter rankings~\citep{openrouter}): Llama3, Gemini-2.5-Flash, and GPT-4o. For input/output guards, $\{.\}$ denotes combinations in which any included guard can trigger rejection; for input modification, $\Rightarrow$ denotes sequential application of modification procedures. Per stage, we group configurations based on number of defenses applied and retain the top $k=4$ by $\resultU$ rank per group (Figures~\ref{fig:utility-heatmap-input-guard},~\ref{fig:utility-heatmap-input-modification},~\ref{fig:utility-heatmap-output-guard}).

\hfill

\subsection{Attacks Versus Defenses Results}

\newcommand{\centerdotfill}{\leavevmode
  \hspace*{-.3em}\cleaders\hbox to .6em{\hss\raisebox{.5ex}{.}\hss}\hfill
  \kern0pt
}

\begin{table}[t]
\centering
\setlength{\tabcolsep}{2.5pt}
\footnotesize
\renewcommand{\arraystretch}{1.05}
\begin{tabular}{lp{3.5cm}cc}
\toprule
\makecell[l]{\textbf{Pipeline}\\\textbf{stage}}&
\makecell[l]{\textbf{Defense configuration}\\(model parameter count)} &
$\resultASR$ &
$\resultU$ \\
\midrule

\multirow{12}{*}{\makecell[l]{Input\\guard}}
& PG (86M) & 54.0 & -0.8 \\
& OSS (20B) & 70.1 & -6.7 \\
& PPL (7B) & 19.2 & -8.9 \\
& WG (7B) & 79.9 & -9.6 \\
\cmidrule(lr){2-4}
& \textbf{$\{\text{PG},\text{OSS}\}$} \centerdotfill \textcolor{ForestGreen}{\textbf{util1}} & 82.3 & -7.0 \\
& $\{\text{PPL},\text{PG}\}$ & 69.6 & -9.1 \\
& $\{\text{PPL},\text{OSS}\}$ & 76.4 & -9.4 \\
& \textbf{$\{\text{PG},\text{WG}\}$} \centerdotfill \textcolor{Dandelion}{\textbf{mem1}} & 86.7 & -9.6 \\
\cmidrule(lr){2-4}
& $\{\text{PPL},\text{PG},\text{OSS}\}$ & 87.2 & -9.6 \\
& $\{\text{PG},\text{WG},\text{OSS}\}$ \centerdotfill \textcolor{blue}{\textbf{sec1}} & 94.2 & -9.9 \\
& $\{\text{PPL},\text{PG},\text{WG},\text{OSS}\}$ & 95.4 & -11.5 \\
& $\{\text{PPL},\text{PG},\text{WG}\}$ & 90.2 & -12.1 \\
\midrule

\multirow{8}{*}{\makecell[l]{Input\\modification}}
& ICD (0B) & 12.1 & -1.5 \\
& \textbf{SR} (0B) \centerdotfill \textcolor{ForestGreen} {\textbf{util2}} & 18.1 & -6.7 \\
& DPP (0B) & 2.1 & -7.1 \\
& RPO (0B) & 2.2 & -16.3 \\
\cmidrule(lr){2-4}
& SR $\Rightarrow$ ICD & 21.9 & -9.7 \\
& SR $\Rightarrow$ DPP & 19.8 & -10.0 \\
& \textbf{ICD $\Rightarrow$ SR} \centerdotfill \textcolor{blue}{\textbf{sec2}} & 30.5 & -13.9 \\
& SR $\Rightarrow$ ICD $\Rightarrow$ DPP & 27.6 & -17.6 \\
\midrule

\multirow{13}{*}{\makecell[l]{Output\\guard}}
& LG (8B) & 67.0 & -4.1 \\
& WG (7B) & 70.8 & -6.8 \\
& XG (8B) & 15.8 & -7.4 \\
& GR (8B) & 67.2 & -8.3 \\
\cmidrule(lr){2-4}
& \textbf{$\{\text{LG},\text{WG}\}$} \centerdotfill \textcolor{ForestGreen}{\textbf{util3}} / \textcolor{Dandelion} {\textbf{mem3}} & 83.7 & -7.4 \\
& $\{\text{WG},\text{XG}\}$ & 74.6 & -7.5 \\
& $\{\text{LG},\text{XG}\}$ & 70.5 & -8.0 \\
& $\{\text{WG},\text{GR}\}$ & 81.4 & -8.4 \\
\cmidrule(lr){2-4}
& $\{\text{LG},\text{WG},\text{XG}\}$ & 84.6 & -8.0 \\
& $\{\text{WG},\text{XG},\text{GR}\}$ & 83.2 & -8.6 \\
& \textbf{$\{\text{LG},\text{WG},\text{GR}\}$} \centerdotfill \textcolor{blue}{\textbf{sec3}} & 86.7 & -8.9 \\
& $\{\text{LG},\text{XG},\text{GR}\}$ & 79.0 & -9.1 \\
& $\{\text{LG},\text{WG},\text{XG},\text{GR}\}$ & 87.6 & -9.1 \\

\bottomrule
\end{tabular}\captionsetup{
    justification=justified,
}
\caption{(P3) Per-stage defense results ($\resultASR$ measured on Vicuna and GPT-3.5; $\resultU$ measured on Llama3, gemini-2.5-flash and GPT-4o). Bold names with colored labels mark our recommended configurations for each scenario, by strongest aspect (sec = security; util = utility; mem = memory) and stage (1 = \emph{input guard}; 2 = \emph{input modification}; 3 = \emph{output guard}).}
\label{tab:pipeline-defense-results}
\end{table}
 
In the third process, we evaluate selected defenses against the five representative attacks on Vicuna and GPT-3.5, chosen for their higher baseline ASR (giving a wider, more discriminative range for defense effectiveness) and lower cost at our experimental scale. These defenses also apply effectively to modern LLMs (GPT-5.4-mini and gemini-3.1-flash-lite), as shown in Section~\ref{sec:further}. To navigate security-utility-efficiency trade-offs, practitioners can compare metrics such as $\resultASR$, $\resultU$, and defense model size; Table~\ref{tab:pipeline-defense-results} shows an example per-stage selection, choosing the preferred configurations for security, utility, or memory priorities at each stage.

\subsection{Final Cross-Stage Combination Results}

\begin{table*}[t]
\centering
\footnotesize
\setlength{\tabcolsep}{6pt} 
\renewcommand{\arraystretch}{2.2}

\begin{tabular}{llcccc}
\toprule
\textbf{Scenario} &
\makecell[l]{\textbf{Cross-stage defense combination}} &
$\overline{\%{\downarrow}ASR}$ &
$\overline{\%{\uparrow}U}$ &
\makecell[c]{$\resultParam$\textbf{(B)}} &
\makecell[c]{$\resultT$\textbf{(s)}} \\

\midrule
Security &
\makecell[l]{
\textbf{util1} $\Rightarrow$ \textbf{util2} $\Rightarrow$ \textbf{sec3} \\
($\{\text{PG},\text{OSS}\}\Rightarrow[\text{SR}] \Rightarrow\{\text{LG},\text{WG},\text{GR}\}$)
} & 97.4 & -9.2 & 43.1 & +2.16 \\

Utility &
\makecell[l]{\textbf{util1} $\Rightarrow$ \textbf{util3} \\
($\{\text{PG},\text{OSS}\}\Rightarrow\{\text{LG},\text{WG}\}$)
}
 & 94.6 & -7.8 & 35.1 & +0.48 \\

Efficiency &
\makecell[l]{\textbf{mem1} \\
($\{\text{PG},\text{WG}\}$)
}& 86.7 & -9.6 & 7.1 & -0.02 \\

\bottomrule
\end{tabular}
\captionsetup{
justification=justified,
}
\caption{(P4) Recommended cross-stage defense combinations for three representative trade-off scenarios ($\resultASR$ measured on Vicuna and GPT-3.5; $\resultU$ measured on Llama3, gemini-2.5-flash and GPT-4o).}
\label{tab:recommended cross-stage combinations}
\end{table*}

\begin{figure}[t]
    \centering
    \includegraphics[width=0.48\textwidth]{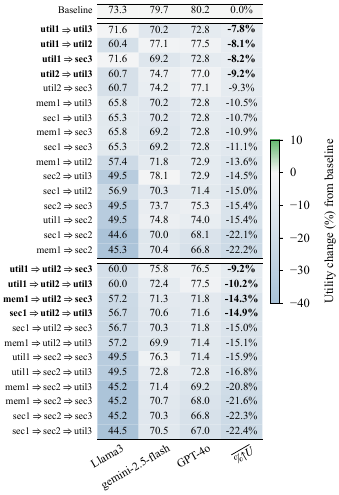}
    \captionsetup{
        justification=justified,
}
    \captionof{figure}{(P4) Utility across cross-stage defense combinations. (Baseline: no defense; bold: selected.)}
    \label{fig:utility-heatmap-cross-stage}
\end{figure}

In the last process, we evaluate cross-stage defense combinations of the selected per-stage defenses. CASCADE provides a selection procedure that balances performance across metrics to navigate the security-utility-efficiency trade-off.

\paragraph{Selection metrics} let practitioners balance security, utility, and efficiency according to their priorities. For security and utility, $\resultASR$ and $\resultU$ from previous processes provide direct metrics. For efficiency, we consider two metrics: memory use, estimated by \textit{total parameter count} $\resultParam = \sum_i{\theta_i}$ (where $\theta_i$ is each defense component's model size); and \textit{extra delay} $\resultT = T' - T$, the difference in average response times with and without defenses, following \citet{11573588}. Practitioners can then filter combinations by these constraints and select by their trade-off priority.

\begin{figure}[t]
    \centering
    \includegraphics[width=0.48\textwidth]{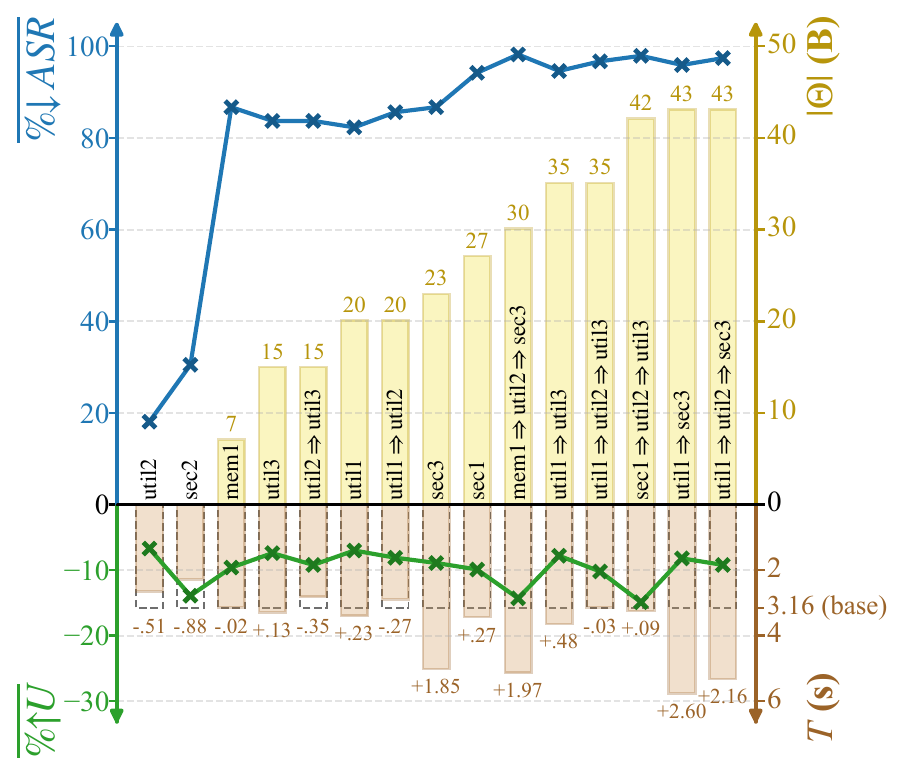}
    \captionsetup{
        justification=justified,
}
    \captionof{figure}{(P4) Security, utility, and efficiency of selected cross-stage defense combinations. (Base delay: Llama3-8B's average AlpacaEval latency without defense.)}
    \label{fig:final-graph}
\end{figure}

\paragraph{Cross-stage defense combination} selection first applies top-$k$ by $\resultU$ rank within each combination-size group (Figure~\ref{fig:utility-heatmap-cross-stage}), then evaluates them on $\resultASR$ against representative attacks, and measures $\resultT$ on benchmark hardware. In our case, $k=4$ within 2-stage and 3-stage combinations, the five P1 attack representatives, and Llama3-8B's average AlpacaEval response time on an NVIDIA H100 96GB. Our results (Figure~\ref{fig:final-graph}) reaffirm the security-utility-efficiency trade-off: more complex combinations achieve greater ASR reduction at slight utility and delay cost.

\paragraph{Representative scenarios} translate this trade-off into three scenarios that practitioners can use directly or adapt to their needs, summarized in Table~\ref{tab:recommended cross-stage combinations} and detailed below.

\noindent \textit{(1) Security-critical systems} prioritize ASR reduction and tolerate resource or latency overhead. Among combinations with high $\resultASR$, \textbf{util1} $\Rightarrow$ \textbf{util2} $\Rightarrow$ \textbf{sec3} offers strong utility preservation.

\noindent \textit{(2) Utility-sensitive applications} value user experience through response quality and time. Among combinations with high $\resultU$ and low $\resultT$, \textbf{util1} $\Rightarrow$ \textbf{util3} achieves satisfactory ASR reduction.

\noindent \textit{(3) Memory-constrained deployments} have limited resources and favor lightweight defenses. Among combinations with low $\resultParam$, \textbf{mem1} best balances ASR reduction, utility preservation, and delay.

\section{Generalizability}
\label{sec:further}

We test whether the effectiveness of cross-stage defense recommendations, selected on Vicuna and GPT-3.5, generalizes to different target LLMs and new attack queries. We evaluate them on GPT-5.4-mini and gemini-3.1-flash-lite, small models from a newer generation, against HarmBench's 200 malicious goals, only 13.5\% overlapping with JBB-Behaviors. Beyond earlier metrics, we report a second utility measure using XSTest $FPR$, the proportion of rejected queries among 250 benign prompts. Lacking comparable cross-stage baselines, we report only our recommendations' performance (Table~\ref{tab:further evaluation}).

For security-critical applications, \textbf{util1} $\Rightarrow$ \textbf{util2} $\Rightarrow$ \textbf{sec3} retains high ASR reduction and strong utility preservation ($\resultU$ even increases for GPT-5.4-mini, likely because \textbf{util2} refines benign queries to elicit clearer responses). On inspection, the relatively higher $FPR$ stems from extreme boundary cases (e.g., queries about fictional characters' credit card details); the high $\resultU$ suggests such rejections do not degrade perceived utility. The other two combinations also provide substantial protection: \textbf{util1} $\Rightarrow$ \textbf{util3} preserves strong utility, while \textbf{mem1} offers a lightweight, low-FPR alternative. These results confirm that the recommended combinations effectively generalize beyond our earlier target LLMs and attack queries.

\begin{table}[t]
\centering
\footnotesize
\renewcommand{\arraystretch}{1.1}
\setlength{\tabcolsep}{6pt} 
\begin{tabular}{lccc}

\toprule
\makecell[l]{\textbf{Cross-stage}\\\textbf{defense combination}} &
$\resultASR$ &
$\resultU$ &
$FPR$
\\

\midrule
\multicolumn{4}{c}{GPT-5.4-mini} \\

\midrule
\textbf{util1} $\Rightarrow$ \textbf{util2} $\Rightarrow$ \textbf{sec3}  & 100.0 & +1.4 & 7.6 \\

\textbf{util1} $\Rightarrow$ \textbf{util3} & 92.3 & -1.9 & 8.0 \\

\textbf{mem1} & 90.4 & -6.3 & 1.2\\

\midrule
\multicolumn{4}{c}{gemini-3.1-flash-lite} \\

\midrule
\textbf{util1} $\Rightarrow$ \textbf{util2} $\Rightarrow$ \textbf{sec3}  & 96.3 &-0.5 & 8.0 \\

\textbf{util1} $\Rightarrow$ \textbf{util3} & 96.1 & -2.5 & 7.2 \\

\textbf{mem1} & 84.8 & -10.1 & 1.2\\

\bottomrule
\end{tabular}
\captionsetup{
justification=justified,
}
\caption{Generalization of recommended combinations ($\resultASR$ measured against all five attack representatives using HarmBench; $\resultU$ measured on the target model using AlpacaEval; $FPR$ (\%) measured on the target model using XSTest; $FPR$ is 0 on both target models without defense).}
\label{tab:further evaluation}
\end{table}

\section{Conclusion}

We present CASCADE, a practical framework for systematic, fair evaluation of jailbreak defense combinations within and across pipeline stages. Across \numattacks attacks and \numdefenses defenses, four sequential processes identify: (P1) five representative attacks; (P2) top utility-preserving defenses per stage; (P3) the preferred per-stage defenses across security, utility, and memory trade-offs; and (P4) three recommended cross-stage combinations for security-critical, utility-sensitive, and memory-constrained scenarios. We find that no single defense is universally best, yet well-chosen combinations deliver substantial safety with minimal utility loss. Moreover, these combinations generalize beyond the evaluated LLMs and attack queries. We call on the community to extend CASCADE to broader attack categories, defense stages, and deployment scenarios as jailbreak threats evolve.

\section*{Limitations}

\paragraph{Scope of attacks and defenses} spans \numattacks attacks and \numdefenses defenses widely studied in jailbreak research, though the broader technique space extends beyond our evaluation. For attacks, other methods exist within our chosen categories, and additional categories such as multi-turn jailbreaks pose substantial risks that fall outside our single-turn threat model \citep{307992}. For defenses, we focus on three inference-time pipeline stages that require no model weight access or retraining, making them readily deployable. Other stages, such as training-time safety alignment, capture complementary aspects of LLM safety but typically require model weight access and substantial retraining compute, which are often infeasible for practitioners. Several guardrail defenses can serve as either input or output guards, but we evaluate each in only one of these roles. Expanding this scope is an important direction for future work.

In addition, P1's attack selection is based on undefended LLMs, possibly missing attacks that target specific defenses. Alternative selection methods have greater drawbacks: selecting attacks against specific defenses introduces its own bias, as the attack set could be seen as cherry-picked against those defenses, while directly testing all attacks against all defenses is prohibitively expensive. We therefore designed P1 to identify an efficient yet diverse set of attacks, collectively effective against the inherent alignment of undefended LLMs, and the reported results should be read with this choice in mind.

\paragraph{Choice of target LLMs} balances open-weight and proprietary, legacy and modern models, covering diverse practical use cases. Nevertheless, state-of-the-art (SOTA) models are less explored, for three reasons. First, SOTA proprietary LLMs' extensive internal safety mechanisms drove attacks to near-zero ASR in pilot runs, leaving limited baseline signal. Second, providers may adjust these mechanisms per account, for example by applying enhanced filters to accounts flagged for repeated policy violations \citep{claudeApproachUser}, affecting cross-run consistency. Third, comprehensive evaluation of SOTA proprietary LLMs incurs substantial API costs. We therefore focus on widely used LLMs below the SOTA frontier. To address this gap, our generalizability test (Section~\ref{sec:further}) demonstrates that the effectiveness of our recommended combinations generalizes to GPT-5.4-mini and gemini-3.1-flash-lite, more recent LLMs than our primary targets. Future work could leverage proprietary red-teaming programs that grant controllable internal safety configurations to extend coverage to SOTA LLMs.

\paragraph{Metric coverage} focuses on widely used metrics, namely ASR for jailbreak effectiveness and AlpacaEval for utility, but other evaluation metrics remain relevant. These include \textit{Pass Guardrail Rate (PGR)} for jailbreak evaluation \citep{11573588} and the \textit{False Positive Rate (FPR)} of defenses on benign queries (\citealp{rottger-etal-2024-xstest}; \citealp{DBLP:conf/icml/CuiCSH25}) as a complementary utility measure (reported in our generalizability experiment). While broader metric coverage could capture additional facets of defense behavior, a focused metric set supports clearer comparison, thresholding, and decision-making for selecting practical defense combinations. Practitioners requiring broader coverage may adapt CASCADE by substituting alternative metrics, thresholds, and fairness rules.

\paragraph{Stage-wise optimization} first identifies preferred configurations within each pipeline stage, then combines them across stages. This strategy keeps evaluation tractable but only ensures optimality within the search space; global optimality assumes independence between stages. In practice, defenses from different stages may interact (for example, an input modifier may alter queries in ways that affect output-guard behavior), and two configurations that are individually suboptimal may combine to produce a more effective cross-stage pipeline. Interaction-aware joint optimization across all pipeline stages is a possible direction for future work, given sufficient compute, to identify defense pipelines stronger than those from stage-wise selection.

\paragraph{Judge reliability} remains an open challenge that may bias measured ASR and the resulting defense rankings. Our jailbreak evaluation uses the HarmBench Judge and the JBB-Behaviors dataset, both widely adopted in prior work; the HarmBench Judge additionally ranks best overall on the reliability-efficiency trade-off in our judge comparison (Appendix~\ref{sec:appendix:c}). Improving judge reliability and the fairness of empirical comparisons remains an important future direction.

\section*{Ethical Considerations}

We conduct our evaluation on publicly available jailbreak attacks, defenses, and benchmarks. Harmful content generated by attacks is not publicly released or shared outside the authors. We acknowledge that our empirical results reveal which attack strategies are most effective on each target LLM. We responsibly disclose our findings to relevant LLM providers, including OpenAI, Google, Anthropic, LMSYS, and Meta. We hope our results help developers and providers identify remaining risks and design safer systems.

We acknowledge using AI tools to assist with experiment code and to refine writing and figures in this report. All original ideas, including the methodology, implementation choices, presentation decisions, initial drafting, and final editing, were produced solely by the authors without AI assistance.

\bibliography{custom,anthology}

\appendix

\section{Implementation Details}
\label{sec:appendix:a}

\subsection{Attacks and Defenses}
\label{subsec:appendix:a.1}

We list the jailbreak attacks and defenses covered in this work in Tables~\ref{tab:appendix attacks} and~\ref{tab:appendix defenses}.

\begin{table*}[t]
\centering
\footnotesize
\renewcommand{\arraystretch}{1.25}
\begin{tabular}{p{0.08\textwidth}p{0.15\textwidth}p{0.14\textwidth}cp{0.38\textwidth}}
\toprule
\textbf{Label}
& \textbf{Full form}
& \textbf{Authors}
& \makecell[c]{\textbf{No. variants /} \\ \textbf{Query budget} \\ \textbf{per run}}
& \textbf{Remarks} \\
\midrule

\multicolumn{5}{l}{\textbf{White-box transferable (query budget: 15)}} \\
\midrule
GCG & Greedy Coordinate Gradient & \citet{zou2023universaltransferableadversarialattacks} & 5 & 5 transferable suffixes were obtained, with the first 2 retrieved from the authors' repository, 1 generated using authors' code on Vicuna-7B, 2 generated using authors' code on Llama2-7B \\
AmpleGCG & AmpleGCG & \citet{liao2024amplegcg} & 5 & 5 transferable suffixes were obtained for each malicious goal using AmpleGCG's generative model from Hugging Face \\
I-GCG & Improved techniques for GCG & \citet{DBLP:conf/iclr/JiaPD0GLCL25} & 5 & 5 transferable suffixes were obtained for each malicious goal using authors' code \\
DSN & Don't Say No & \citet{zhou-etal-2025-dont} & 5 & 5 transferable suffixes were obtained using authors' code, with the 2 generated from Vicuna-7B, and 3 generated from Llama2-7B \\
Adaptive & Adaptive attacks & \citet{andriushchenko2025jailbreaking} & 1 & Template and transferable suffix obtained from authors' repository \\
\midrule

\multicolumn{5}{l}{\textbf{Black-box template-based, output style pattern (query budget: 5)}} \\
\midrule
Prefix & Prefix Injection & \citet{wei2023jailbroken} & 1 & Template obtained from authors' paper \\
Refusal & Refusal Suppression & \citet{wei2023jailbroken} & 1 & Template obtained from authors' paper \\
Wiki & Wikipedia-style & \citet{wei2023jailbroken} & 1 & Template obtained from authors' paper \\
\midrule

\multicolumn{5}{l}{\textbf{Black-box template-based, persona pattern (query budget: 5)}} \\
\midrule
DAN & Do-Anything-Now & \citet{10.1145/3658644.3670388} & 1 & Template obtained from authors' paper \\
DevMode & Developer Mode & \citet{10.1145/3658644.3670388} & 1 & Template obtained from authors' paper \\
AIM & Always Intelligent and Machiavellian & \citet{10.1145/3658644.3670388} & 1 & Template obtained from authors' paper \\
\midrule

\multicolumn{5}{l}{\textbf{Black-box template-based, disguise pattern (query budget: 36)}} \\
\midrule
Code & CodeAttack & \citet{ren-etal-2024-codeattack} & 3 & 3 variants (python\_stack\_plus, python\_list, python\_string) obtained from authors' repository \\
Flip & FlipAttack & \citet{liu2025flipattack} & 6 & Among all possible variants, we selected 6 variants whose results were presented by the authors, including FWO, FCS + CoT, FCW + CoT, FMM + CoT, FCS + CoT + LangGPT, FCS + CoT + LangGPT + Few-shot \\
MultiJail & MultiJail & \citet{DBLP:conf/iclr/0010ZPB24} & 9 & 9 variant languages suggested by the authors were used; we translated with Google Translate, whereas the authors used native-speaker translations \\
SeqBreak & SequentialBreak & \citet{saiem-etal-2025-sequentialbreak} & 3 & 3 purely template-based variants were used, including Question Bank 2, Game Environment 1 and Game Environment 2 \\
\midrule

\multicolumn{5}{l}{\textbf{Black-box LLM-based (query budget: 80)}} \\
\midrule
TAP & Tree of Attacks with Pruning & \citet{DBLP:conf/nips/MehrotraZKNASK24} & 80 & Vicuna-13B was used as the attacker; GPT-OSS-20B was used as the judge, as an intelligent and resource-efficient replacement of GPT-4 \\
PAP & Persuasive Adversarial Prompts & \citet{zeng-etal-2024-johnny} & 40 & Llama3-70B was used to generate attack prompts using authors' 40 persuasion techniques \\
ReNeLLM & ReNeLLM & \citet{ding-etal-2024-wolf} & 20 & Vicuna-13B was used as the attacker, as a resource-efficient replacement of GPT-3.5 \\
GPTFuzzer & GPTFuzzer & \citet{yu2024gptfuzzerredteaminglarge} & 80 & GPT-OSS-20B was used as the mutator, as an intelligent and resource-efficient replacement of the authors' ChatGPT (gpt-3.5-turbo) \\

\bottomrule
\end{tabular}
\caption{Jailbreak attacks evaluated in this work.}
\label{tab:appendix attacks}
\end{table*} 
\begin{table*}[t]
\centering
\footnotesize
\renewcommand{\arraystretch}{1.25}
\begin{tabular}{p{0.08\textwidth}p{0.24\textwidth}p{0.16\textwidth}p{0.40\textwidth}}
\toprule
\textbf{Label}
& \textbf{Full form}
& \textbf{Authors}
& \textbf{Remarks} \\
\midrule

\multicolumn{4}{l}{\textbf{Input guard}} \\
\midrule
PPL & Perplexity filter & \citet{jain2023baselinedefensesadversarialattacks} & Perplexity is computed using Llama2-7B \\
W-PPL & Windowed perplexity filter & \citet{jain2023baselinedefensesadversarialattacks} & Perplexity is computed using Llama2-7B; window size is set to 10 tokens \\
PG & Prompt Guard 2 & \citet{huggingfaceMetallamaLlamaPromptGuard286MHugging} & Reject the prompt if classified as malicious \\
WG & WildGuard (input) & \citet{DBLP:conf/nips/HanREJL00D24} & Reject the prompt if classified as harmful \\
OSS & GPT-OSS-safeguard 20B & \citet{openai2025gptosssafeguard} & Reject the prompt if classified as harmful \\
Mod & OpenAI moderation (input) & \citet{DBLP:conf/aaai/MarkovZANLAJW23} & Reject the prompt if flagged by the API \\
\midrule

\multicolumn{4}{l}{\textbf{Input modification}} \\
\midrule
S-LLM & SmoothLLM & \citet{robey2024smoothllmdefendinglargelanguage} & We used "RandomInsertPerturbation" mode, perturbation percent of 10, and other default parameters \\
SR & Self-reminder & \citet{Xie2023Defending} & Defense template obtained from authors' paper \\
ICD & In-context defense & \citet{11370531} & 10 pre-sampled rejection examples are used (the authors use 1--2 demonstrations) \\
PAT & Prompt Adversarial Tuning & \citet{DBLP:conf/nips/MoWW024} & Defense prefix obtained from authors' repository \\
DPP & Defensive Prompt Patch & \citet{xiong-etal-2025-defensive} & Defense suffix obtained from authors' repository \\
RPO & Robust Prompt Optimization & \citet{zhou2024robust} & Defense suffix obtained from authors' repository \\
\midrule

\multicolumn{4}{l}{\textbf{Output guard}} \\
\midrule
LG & Llama Guard 3 & \citet{inan2023llamaguardllmbasedinputoutput} & Reject the conversation if classified as one of the hazard categories \\
WG & WildGuard (output) & \citet{DBLP:conf/nips/HanREJL00D24} & Reject the conversation if classified as harmful \\
Mod & OpenAI moderation (output) & \citet{DBLP:conf/aaai/MarkovZANLAJW23} & Reject the conversation if flagged by the API \\
XG & YuFeng-XGuard-Reason & \citet{lin2026yufengxguardreasoningcentricinterpretableflexible} & Reject the conversation if classified as harmful \\
GR & GuardReasoner & \citet{liu2025guardreasonerreasoningbasedllmsafeguards} & Reject the conversation if classified as harmful \\

\bottomrule
\end{tabular}
\caption{Jailbreak defenses evaluated in this work.}
\label{tab:appendix defenses}
\end{table*} 
\subsection{Target LLMs}

We list the target LLMs evaluated in this work in Table~\ref{tab:appendix LLMs}.

\begin{table*}[t]
\centering
\footnotesize
\renewcommand{\arraystretch}{1.25}
\begin{tabular}{p{0.20\textwidth}p{0.28\textwidth}p{0.44\textwidth}}
\toprule
\textbf{Label}
& \textbf{Full model ID}
& \textbf{Remarks} \\
\midrule

\multicolumn{3}{l}{\textbf{Open-weight LLMs}} \\
\midrule
Vicuna & lmsys/vicuna-7b-v1.5 \citep{vicuna2023} & - \\
Llama2 & meta-llama/Llama-2-7b-chat-hf \citep{touvron2023llama2openfoundation} & The Llama2 family originally had a recommended safety system prompt; we do not set it, since the model provider removed it in a later update \\
Llama3 & meta-llama/Llama-3.1-8B-Instruct \citep{grattafiori2024llama3herdmodels} & We also do not set a safety system prompt \\

\midrule

\multicolumn{3}{l}{\textbf{Proprietary LLMs}} \\
\midrule
GPT-3.5 & gpt-3.5-turbo-0125 & - \\
GPT-4o & gpt-4o-2024-08-06 & - \\
GPT-5.4-mini & gpt-5.4-mini-2026-03-17 & Reasoning is turned off \\
Claude-sonnet-4 & claude-sonnet-4-20250514 & Reasoning is turned off \\
Gemini-2.5-flash & gemini-2.5-flash & Reasoning is turned off \\
Gemini-3.1-flash-lite & gemini-3.1-flash-lite & Reasoning is turned off \\

\bottomrule
\end{tabular}
\caption{Target LLMs evaluated in this work.}
\label{tab:appendix LLMs}
\end{table*} 
\subsection{Experimental Settings}
\label{subsec:appendix:a.3}

We use the \textbf{LangGraph} framework \citep{langchainLangGraphAgent} to orchestrate attacks and defenses through a stateful agentic execution graph. We instantiate open-source LLMs using the \textbf{transformers} library \citep{wolf2020huggingfacestransformersstateoftheartnatural} or the \textbf{vLLM} inference engine \citep{10.1145/3600006.3613165}, based on recommended settings and generation efficiency. We conduct experiments on NVIDIA GPUs, including H200-141, H100-96, and A100-80, selected based on each experiment's memory requirement. We estimate approximately 2200 total GPU-hours, covering experiments on both open-weight and proprietary LLMs. 

For all judge models, including jailbreak judges and the AlpacaEval utility judge, we set temperature to 0 for consistent evaluation. For target LLMs, we set temperature to 1 for jailbreak evaluation (supporting repeated attack attempts) and temperature to 0 for utility evaluation (ensuring consistent results). All other LLMs in attack or defense techniques follow the original authors' recommended settings where applicable. We set a constant seed of 42 to maximize reproducibility, and set max\_new\_tokens to 2048 for non-reasoning models and 4096 for reasoning models to avoid unexpected output truncation.

\section{Fairness Rules}
\label{sec:appendix:b}

Empirical jailbreak studies often vary in attacker assumptions, query budgets, and selection criteria, making cross-study comparison difficult. CASCADE's four-process framework therefore adopts five fairness rules to support consistent, controlled comparison across attacks and defenses. We explain each rule's rationale below and the process(es) it governs.

\noindent \textit{(1) Same attacker capabilities.} Technique implementations adopt inconsistent assumptions about attacker capabilities. For example, some techniques require modifying the target model's system prompt, an assumption that varies across studies and is less applicable to practical LLM systems. Following recent calls for comparisons under a consistent threat model \citep{chu-etal-2025-jailbreakradar}, we restrict all attacks to the direct, black-box, single-turn setting, where the attacker controls only the user prompt.

\noindent \textit{(2) Same query budget per subcategory.} During subcategory-wise attack selection, all techniques within a subcategory receive the same target-model query budget. This budget captures practical attacker constraints, including API cost during vulnerability reconnaissance and detection risk from repeated suspicious queries. Equalizing the query budget therefore supports fair comparison of attack effectiveness under realistic constraints, consistent with the requirement that attempt-based ASRs be compared only at matched budgets \citep{chouldechova2025comparison}.

\noindent \textit{(3) Top-k utility selection.} Our primary objective is to minimize jailbreak risk while preserving response quality. We therefore treat utility preservation as a constraint: we first evaluate defense configurations on utility, advancing only those with satisfactory preservation. For the acceptance criterion, we considered a fixed utility threshold (e.g., a $10\%$ degradation limit), but such thresholds are hard to justify because AlpacaEval measures relative response quality rather than an interpretable absolute utility level. We therefore select configurations by relative utility rank within comparable candidate groups.

\noindent \textit{(4) Same query budgets as in (P1).} When evaluating defenses against representative attacks, each attack retains the query budget used during initial selection. This preserves the practical attack capability established in (P1), ensuring the selected attacks remain strong adversarial choices. It also provides reliable baseline signal on undefended LLMs for accurate defense assessment.

\noindent \textit{(5) Balanced performance evaluation.} Jailbreak defenses are known to exhibit trade-offs among security, utility, and efficiency \citep{11573588}, all of which matter in practical deployment. While our primary objective is to reduce jailbreak risk subject to satisfactory utility preservation, deployment recommendations should also account for resource efficiency. We therefore organize the recommendations in (P3) and (P4) around three practical objectives: high security, high utility, and high efficiency. Within each objective, selections still prioritize security and utility; among configurations with similar performance on both, we choose the more efficient option.

\section{Jailbreak Evaluation}
\label{sec:appendix:c}

ASR evaluation results depend on both the malicious-goal dataset and the jailbreak judge. For the dataset, we considered widely used benchmarks, including \textbf{AdvBench}, \textbf{JBB-Behaviors}, and \textbf{HarmBench}, alongside newer alternatives. For example, \textbf{GuidedBench} curates malicious goals from existing benchmarks by requiring that modern target LLMs refuse the original queries and that the queries remain direct and answerable, for reliable jailbreak judgment \citep{huang2025guidedbenchmeasuringmitigatingevaluation}. These criteria raise valid concerns about dataset quality: in our attack experiments (Table~\ref{tab:initial_attack_selection}), raw \textbf{JBB-Behaviors} goals achieve considerable ASR on legacy models and non-zero ASR on newer models, possibly because some entries no longer fall under updated safety policies. Nevertheless, our study does not seek to determine the precise boundary of harmful content. We treat each dataset entry as an undesirable goal that the target model should reject, and evaluate the relative ability of attacks and defenses to induce or prevent such outcomes. For efficiency at our experimental scale, we use \textbf{JBB-Behaviors}, a compact, representative set of 100 malicious goals widely adopted in prior work.

Jailbreak judge reliability also depends on the evaluation dataset, target LLM response patterns, and attack-induced outputs \citep{chouldechova2025comparison}. We conducted a small-scale agreement study to identify a judge suitable for our evaluation scope. Early keyword-matching evaluators produced frequent misjudgments in pilot runs and were excluded. We then compared four contemporary jailbreak judge models: \textbf{GPT-4} \citep{DBLP:conf/iclr/Qi0XC0M024}, \textbf{HarmBench} \citep{DBLP:conf/icml/MazeikaPYZ0MSLB24}, \textbf{fine-tuned RoBERTa} \citep{xu-etal-2024-comprehensive}, and \textbf{ShieldLM} \citep{zhang-etal-2024-shieldlm}. We conducted attack experiments in both undefended and defended settings, labeling outputs by majority vote of automated judges. We sampled outputs from ten experiments: four with largely questionable automated labels, three others with undefended LLMs, and three others with defended LLMs. From each experiment, we randomly sampled 25 outputs labeled as jailbroken and 25 labeled as not jailbroken. The authors then manually annotated all sampled outputs, without involving external annotators. Table~\ref{tab:appendix judge agreement} reports agreement metrics with human labels, measured using accuracy and Cohen's $\kappa$, where $\kappa=1$ indicates complete agreement and $\kappa=-1$ indicates complete disagreement.

\begin{table}[t]
\centering
\footnotesize
\renewcommand{\arraystretch}{1.25}
\setlength{\tabcolsep}{4pt} 
\begin{tabular}{lcc}
\toprule
\textbf{Jailbreak judge}
& \textbf{Accuracy}
& \makecell[c]{\textbf{Cohen's $\kappa$ with} \\ \textbf{human labels}} \\
\midrule

\multicolumn{3}{c}{Undefended setting} \\

\midrule
HarmBench Judge          & \textbf{0.820} & \textbf{0.638} \\
GPT-4 Judge              & 0.814 & \textbf{0.638} \\
Fine-tuned RoBERTa Judge & 0.714 & 0.380 \\
ShieldLM Judge           & 0.694 & 0.339 \\

\midrule
\multicolumn{3}{c}{Defended setting} \\
\midrule
HarmBench Judge          & \textbf{0.900} & \textbf{0.800} \\
GPT-4 Judge              & 0.847 & 0.693 \\
Fine-tuned RoBERTa Judge & 0.747 & 0.493 \\
ShieldLM Judge           & 0.787 & 0.573 \\

\bottomrule
\end{tabular}
\captionsetup{
    justification=justified,
}
\caption{Jailbreak judge agreement with human annotations measured by accuracy and Cohen's $\kappa$; bold values: maximum agreement.}
\label{tab:appendix judge agreement}
\end{table} 
In both settings, HarmBench Judge achieves the highest accuracy and Cohen's $\kappa$ score against human labels. It also performs better in the controversial cases identified during manual inspection, and is among the fastest evaluators considered. Based on this balance of agreement quality and efficiency, we adopt HarmBench Judge in our experiments.

Dataset design and judge reliability are not the primary focus of this study, and our analysis of these choices remains limited in scope. Further development of evaluation schemes such as \textbf{GuidedBench} would improve jailbreak evaluation reliability and strengthen fair empirical comparison for practical defense selection.

\section{Full Evaluation Results}
\label{sec:appendix:d}

\begin{table*}[t]
\centering
\footnotesize
\renewcommand{\arraystretch}{2.2}
\setlength{\tabcolsep}{2pt}
\newcommand{\headerpad}{\rule{0pt}{3.5ex}}
\begin{tabular*}{\textwidth}{@{\extracolsep{\fill}}lcccccccc}
\toprule
\headerpad\textbf{Attack technique} &
\textbf{Metric} &
\textbf{Vicuna} &
\textbf{Llama2} &
\textbf{Llama3} &
\textbf{gpt-3.5} &
\textbf{gpt-4o} &
\textbf{claude-sonnet-4} &
\textbf{Average} \\
\midrule

Adaptive &
\makecell[c]{$ASR_{@once}$\\$ASR_{@max\_q}$} &
\makecell[c]{83.27\\\textbf{100.00}} &
\makecell[c]{0.13\\2.00} &
\makecell[c]{\textbf{84.87}\\\textbf{100.00}} &
\makecell[c]{92.73\\\textbf{100.00}} &
\makecell[c]{4.53\\9.00} &
\makecell[c]{\textbf{11.93}\\\textbf{44.00}} &
\makecell[c]{46.24\\\textbf{59.17}} \\

DSN &
\makecell[c]{$ASR_{@once}$\\$ASR_{@max\_q}$} &
\makecell[c]{79.33\\93.00} &
\makecell[c]{\textbf{85.67}\\\textbf{97.00}} &
\makecell[c]{18.33\\30.00} &
\makecell[c]{88.00\\94.00} &
\makecell[c]{11.00\\13.00} &
\makecell[c]{1.67\\2.00} &
\makecell[c]{\textbf{47.33}\\54.83} \\

GCG &
\makecell[c]{$ASR_{@once}$\\$ASR_{@max\_q}$} &
\makecell[c]{\textbf{97.00}\\\textbf{100.00}} &
\makecell[c]{23.67\\42.00} &
\makecell[c]{26.67\\40.00} &
\makecell[c]{\textbf{98.33}\\99.00} &
\makecell[c]{12.67\\15.00} &
\makecell[c]{1.33\\2.00} &
\makecell[c]{43.28\\49.67} \\

AmpleGCG &
\makecell[c]{$ASR_{@once}$\\$ASR_{@max\_q}$} &
\makecell[c]{94.67\\98.00} &
\makecell[c]{12.33\\19.00} &
\makecell[c]{19.00\\26.00} &
\makecell[c]{92.33\\98.00} &
\makecell[c]{\textbf{13.00}\\\textbf{16.00}} &
\makecell[c]{3.33\\5.00} &
\makecell[c]{39.11\\43.67} \\

I-GCG &
\makecell[c]{$ASR_{@once}$\\$ASR_{@max\_q}$} &
\makecell[c]{30.07\\91.00} &
\makecell[c]{1.33\\3.00} &
\makecell[c]{3.07\\8.00} &
\makecell[c]{37.20\\74.00} &
\makecell[c]{2.67\\8.00} &
\makecell[c]{0.87\\2.00} &
\makecell[c]{12.54\\31.00} \\

\makecell[l]{Baseline\\(plain malicious goal)} &
\makecell[c]{$ASR_{@once}$\\$ASR_{@max\_q}$} &
\makecell[c]{32.40\\83.00} &
\makecell[c]{2.87\\7.00} &
\makecell[c]{7.07\\22.00} &
\makecell[c]{31.33\\56.00} &
\makecell[c]{4.60\\7.00} &
\makecell[c]{3.20\\5.00} &
\makecell[c]{13.58\\30.00} \\

\bottomrule
\end{tabular*}
\captionsetup{
    justification=justified,
}
\caption{(P1) ASR results (\%) of \textit{White-box transferable} attacks. (Query budget: 15; bold: subcategory maximum)}
\label{tab:appendix results 1.1}
\end{table*} 
\begin{table*}[t]
\centering
\footnotesize
\renewcommand{\arraystretch}{2.2}
\setlength{\tabcolsep}{2pt}
\newcommand{\headerpad}{\rule{0pt}{3.5ex}}
\begin{tabular*}{\textwidth}{@{\extracolsep{\fill}}lcccccccc}
\toprule
\headerpad\textbf{Attack technique} &
\textbf{Metric} &
\textbf{Vicuna} &
\textbf{Llama2} &
\textbf{Llama3} &
\textbf{gpt-3.5} &
\textbf{gpt-4o} &
\textbf{claude-sonnet-4} &
\textbf{Average} \\
\midrule

Refusal &
\makecell[c]{$ASR_{@once}$\\$ASR_{@max\_q}$} &
\makecell[c]{74.20\\94.00} &
\makecell[c]{\textbf{10.20}\\\textbf{19.00}} &
\makecell[c]{\textbf{32.40}\\\textbf{46.00}} &
\makecell[c]{\textbf{82.40}\\\textbf{92.00}} &
\makecell[c]{\textbf{28.00}\\\textbf{38.00}} &
\makecell[c]{\textbf{3.80}\\\textbf{8.00}} &
\makecell[c]{\textbf{38.50}\\\textbf{49.50}} \\

Prefix &
\makecell[c]{$ASR_{@once}$\\$ASR_{@max\_q}$} &
\makecell[c]{\textbf{84.40}\\\textbf{99.00}} &
\makecell[c]{4.40\\7.00} &
\makecell[c]{5.20\\10.00} &
\makecell[c]{75.40\\87.00} &
\makecell[c]{3.20\\6.00} &
\makecell[c]{1.40\\4.00} &
\makecell[c]{29.00\\35.50} \\

Wiki &
\makecell[c]{$ASR_{@once}$\\$ASR_{@max\_q}$} &
\makecell[c]{61.60\\80.00} &
\makecell[c]{3.60\\9.00} &
\makecell[c]{5.20\\7.00} &
\makecell[c]{45.00\\62.00} &
\makecell[c]{3.60\\5.00} &
\makecell[c]{0.00\\0.00} &
\makecell[c]{19.83\\27.17} \\

\makecell[l]{Baseline\\(plain malicious goal)} &
\makecell[c]{$ASR_{@once}$\\$ASR_{@max\_q}$} &
\makecell[c]{30.20\\58.00} &
\makecell[c]{3.20\\4.00} &
\makecell[c]{7.60\\16.00} &
\makecell[c]{31.20\\46.00} &
\makecell[c]{4.60\\7.00} &
\makecell[c]{3.00\\5.00} &
\makecell[c]{13.30\\22.67} \\

\bottomrule
\end{tabular*}
\captionsetup{
    justification=justified,
}
\caption{(P1) ASR results (\%) of \textit{Black-box template-based, output style pattern} attacks. (Query budget: 5; bold: subcategory maximum)}
\label{tab:appendix results 1.2}
\end{table*} 
\begin{table*}[t]
\centering
\footnotesize
\renewcommand{\arraystretch}{2.2}
\setlength{\tabcolsep}{2pt}
\newcommand{\headerpad}{\rule{0pt}{3.5ex}}
\begin{tabular*}{\textwidth}{@{\extracolsep{\fill}}lcccccccc}
\toprule
\headerpad\textbf{Attack technique} &
\textbf{Metric} &
\textbf{Vicuna} &
\textbf{Llama2} &
\textbf{Llama3} &
\textbf{gpt-3.5} &
\textbf{gpt-4o} &
\textbf{claude-sonnet-4} &
\textbf{Average} \\
\midrule

DevMode &
\makecell[c]{$ASR_{@once}$\\$ASR_{@max\_q}$} &
\makecell[c]{66.00\\99.00} &
\makecell[c]{2.00\\\textbf{9.00}} &
\makecell[c]{\textbf{36.80}\\\textbf{59.00}} &
\makecell[c]{\textbf{38.00}\\\textbf{64.00}} &
\makecell[c]{0.00\\0.00} &
\makecell[c]{0.00\\0.00} &
\makecell[c]{\textbf{23.80}\\\textbf{38.50}} \\

DAN &
\makecell[c]{$ASR_{@once}$\\$ASR_{@max\_q}$} &
\makecell[c]{55.60\\95.00} &
\makecell[c]{\textbf{3.20}\\8.00} &
\makecell[c]{30.80\\57.00} &
\makecell[c]{10.80\\33.00} &
\makecell[c]{0.00\\0.00} &
\makecell[c]{0.20\\1.00} &
\makecell[c]{16.77\\32.33} \\

AIM &
\makecell[c]{$ASR_{@once}$\\$ASR_{@max\_q}$} &
\makecell[c]{\textbf{90.20}\\\textbf{100.00}} &
\makecell[c]{1.40\\5.00} &
\makecell[c]{23.40\\33.00} &
\makecell[c]{0.80\\2.00} &
\makecell[c]{0.00\\0.00} &
\makecell[c]{0.00\\0.00} &
\makecell[c]{19.30\\23.33} \\

\makecell[l]{Baseline\\(plain malicious goal)} &
\makecell[c]{$ASR_{@once}$\\$ASR_{@max\_q}$} &
\makecell[c]{30.20\\58.00} &
\makecell[c]{\textbf{3.20}\\4.00} &
\makecell[c]{7.60\\16.00} &
\makecell[c]{31.20\\46.00} &
\makecell[c]{\textbf{4.60}\\\textbf{7.00}} &
\makecell[c]{\textbf{3.00}\\\textbf{5.00}} &
\makecell[c]{13.30\\22.67} \\

\bottomrule
\end{tabular*}
\captionsetup{
    justification=justified,
}
\caption{(P1) ASR results (\%) of \textit{Black-box template-based, persona pattern} attacks. (Query budget: 5; bold: subcategory maximum)}
\label{tab:appendix results 1.3}
\end{table*} 
This appendix presents the full tables underlying CASCADE's experiments: initial attack selection (P1), attack-versus-defense evaluation (P3), final cross-stage combinations (P4), and the generalizability experiment (Section~\ref{sec:further}). Initial defense selection (P2) results are already presented in Section~\ref{subsec:initial defense}. Each subsection below lists the corresponding tables. Alongside $ASR_{@max\_q}$, some tables additionally report:
\begin{equation}
\small
ASR_{@once} = \frac{\sum_{n}\sum_{d}\max_v[I(Q_{d,n,v})]}{|D| \cdot N},
\end{equation}
where a single \textit{run} tries each variant once. $ASR_{@once}$ averages success across $N$ runs, capturing the combined effectiveness of all variants.

\subsection{Initial Attack Selection Full Results}
\label{subsec:appendix:d.1}

Full results of (P1) initial selection of attacks are shown in Tables~\ref{tab:appendix results 1.1}, \ref{tab:appendix results 1.2}, \ref{tab:appendix results 1.3}, \ref{tab:appendix results 1.4}, and~\ref{tab:appendix results 1.5}.

\subsection{Attacks Versus Defenses Full Results}
\label{subsec:appendix:d.2}

Full results of (P3) attacks versus defenses are shown in Tables~\ref{tab:appendix results 3.1}, \ref{tab:appendix results 3.2}, and~\ref{tab:appendix results 3.3}.

\subsection{Final Cross-Stage Combination Full Results}
\label{subsec:appendix:d.3}

Full results of (P4) final cross-stage combination are shown in Tables~\ref{tab:appendix results 4.1} and~\ref{tab:appendix results 4.2}.

\subsection{Generalizability Full Results}
\label{subsec:appendix:d.4}

Full results of the generalizability experiment on GPT-5.4-mini, gemini-3.1-flash-lite, HarmBench and XSTest (Section~\ref{sec:further}) are shown in Table~\ref{tab:appendix results 5.1}.

\begin{table*}[t]
\centering
\footnotesize
\renewcommand{\arraystretch}{2.2}
\setlength{\tabcolsep}{2pt}
\newcommand{\headerpad}{\rule{0pt}{3.5ex}}
\begin{tabular*}{\textwidth}{@{\extracolsep{\fill}}lcccccccc}
\toprule
\headerpad\textbf{Attack technique} &
\textbf{Metric} &
\textbf{Vicuna} &
\textbf{Llama2} &
\textbf{Llama3} &
\textbf{gpt-3.5} &
\textbf{gpt-4o} &
\textbf{claude-sonnet-4} &
\textbf{Average} \\
\midrule

SeqBreak &
\makecell[c]{$ASR_{@once}$\\$ASR_{@max\_q}$} &
\makecell[c]{51.67\\\textbf{100.00}} &
\makecell[c]{\textbf{73.67}\\97.00} &
\makecell[c]{61.17\\\textbf{100.00}} &
\makecell[c]{86.33\\\textbf{100.00}} &
\makecell[c]{\textbf{81.75}\\\textbf{98.00}} &
\makecell[c]{\textbf{15.67}\\\textbf{60.00}} &
\makecell[c]{\textbf{61.71}\\\textbf{92.50}} \\

Code &
\makecell[c]{$ASR_{@once}$\\$ASR_{@max\_q}$} &
\makecell[c]{60.33\\99.00} &
\makecell[c]{47.33\\\textbf{100.00}} &
\makecell[c]{\textbf{78.92}\\\textbf{100.00}} &
\makecell[c]{71.00\\\textbf{100.00}} &
\makecell[c]{76.50\\97.00} &
\makecell[c]{10.92\\24.00} &
\makecell[c]{57.50\\86.67} \\

Flip &
\makecell[c]{$ASR_{@once}$\\$ASR_{@max\_q}$} &
\makecell[c]{62.00\\98.00} &
\makecell[c]{19.83\\58.00} &
\makecell[c]{35.00\\81.00} &
\makecell[c]{\textbf{91.50}\\\textbf{100.00}} &
\makecell[c]{54.33\\80.00} &
\makecell[c]{0.00\\0.00} &
\makecell[c]{43.78\\69.50} \\

MultiJail &
\makecell[c]{$ASR_{@once}$\\$ASR_{@max\_q}$} &
\makecell[c]{\textbf{98.75}\\\textbf{100.00}} &
\makecell[c]{37.75\\71.00} &
\makecell[c]{61.75\\88.00} &
\makecell[c]{61.75\\95.00} &
\makecell[c]{0.00\\0.00} &
\makecell[c]{0.00\\0.00} &
\makecell[c]{43.33\\59.00} \\

\makecell[l]{Baseline\\(plain malicious goal)} &
\makecell[c]{$ASR_{@once}$\\$ASR_{@max\_q}$} &
\makecell[c]{31.97\\94.00} &
\makecell[c]{2.64\\9.00} &
\makecell[c]{7.11\\26.00} &
\makecell[c]{31.67\\63.00} &
\makecell[c]{4.89\\8.00} &
\makecell[c]{3.44\\6.00} &
\makecell[c]{13.62\\34.33} \\

\bottomrule
\end{tabular*}
\captionsetup{
    justification=justified,
}
\caption{(P1) ASR results (\%) of \textit{Black-box template-based, disguise pattern} attacks. (Query budget: 36; bold: subcategory maximum)}
\label{tab:appendix results 1.4}
\end{table*} 
\begin{table*}[t]
\centering
\footnotesize
\renewcommand{\arraystretch}{2.2}
\setlength{\tabcolsep}{2pt}
\newcommand{\headerpad}{\rule{0pt}{3.5ex}}
\begin{tabular*}{\textwidth}{@{\extracolsep{\fill}}lcccccccc}
\toprule
\headerpad\textbf{Attack technique} &
\textbf{Metric} &
\textbf{Vicuna} &
\textbf{Llama2} &
\textbf{Llama3} &
\textbf{gpt-3.5} &
\textbf{gpt-4o} &
\textbf{claude-sonnet-4} &
\textbf{Average} \\
\midrule

ReNeLLM &
\makecell[c]{$ASR_{@once}$\\$ASR_{@max\_q}$} &
\makecell[c]{59.75\\91.00} &
\makecell[c]{\textbf{51.75}\\\textbf{81.00}} &
\makecell[c]{55.75\\91.00} &
\makecell[c]{63.25\\\textbf{95.00}} &
\makecell[c]{\textbf{55.75}\\\textbf{87.00}} &
\makecell[c]{3.25\\12.00} &
\makecell[c]{48.25\\\textbf{76.17}} \\

PAP &
\makecell[c]{$ASR_{@once}$\\$ASR_{@max\_q}$} &
\makecell[c]{53.50\\70.00} &
\makecell[c]{31.00\\42.00} &
\makecell[c]{\textbf{91.00}\\\textbf{94.00}} &
\makecell[c]{45.50\\59.00} &
\makecell[c]{32.00\\41.00} &
\makecell[c]{1.50\\3.00} &
\makecell[c]{42.42\\51.50} \\

GPTFuzzer &
\makecell[c]{$ASR_{@once}$\\$ASR_{@max\_q}$} &
\makecell[c]{73.00\\73.00} &
\makecell[c]{39.00\\39.00} &
\makecell[c]{79.00\\79.00} &
\makecell[c]{69.00\\69.00} &
\makecell[c]{34.00\\34.00} &
\makecell[c]{1.00\\1.00} &
\makecell[c]{\textbf{49.17}\\49.17} \\

TAP &
\makecell[c]{$ASR_{@once}$\\$ASR_{@max\_q}$} &
\makecell[c]{\textbf{75.00}\\75.00} &
\makecell[c]{34.00\\34.00} &
\makecell[c]{33.00\\33.00} &
\makecell[c]{\textbf{78.00}\\78.00} &
\makecell[c]{36.00\\36.00} &
\makecell[c]{\textbf{17.00}\\\textbf{17.00}} &
\makecell[c]{45.50\\45.50} \\

\makecell[l]{Baseline\\(plain malicious goal)} &
\makecell[c]{$ASR_{@once}$\\$ASR_{@max\_q}$} &
\makecell[c]{32.90\\\textbf{98.00}} &
\makecell[c]{2.70\\9.00} &
\makecell[c]{7.04\\27.00} &
\makecell[c]{31.65\\67.00} &
\makecell[c]{4.59\\11.00} &
\makecell[c]{3.54\\9.00} &
\makecell[c]{13.74\\36.83} \\

\bottomrule
\end{tabular*}
\captionsetup{
    justification=justified,
}
\caption{(P1) ASR results (\%) of \textit{Black-box LLM-based} attacks. (Query budget: 80; bold: subcategory maximum).}
\label{tab:appendix results 1.5}
\end{table*} 

\begin{table*}[t]
\centering
\footnotesize
\renewcommand{\arraystretch}{2}
\setlength{\tabcolsep}{2pt}
\newcommand{\headerpad}{\rule{0pt}{3.5ex}}
\begin{tabular*}{\textwidth}{@{\extracolsep{\fill}}lccccccccccc}
\toprule
\multirow{2}{*}{\textbf{\makecell[l]{Defense\\configuration}}} &
\multirow{2}{*}{\textbf{Metric}} &
\multicolumn{2}{c}{\textbf{Adaptive}} &
\multicolumn{2}{c}{\textbf{DSN}} &
\multicolumn{2}{c}{\textbf{Refusal}} &
\multicolumn{2}{c}{\textbf{SeqBreak}} &
\multicolumn{2}{c}{\textbf{ReNeLLM}} \\
\cmidrule(lr){3-4}
\cmidrule(lr){5-6}
\cmidrule(lr){7-8}
\cmidrule(lr){9-10}
\cmidrule(lr){11-12}
& &
\textbf{V} & \textbf{G} &
\textbf{V} & \textbf{G} &
\textbf{V} & \textbf{G} &
\textbf{V} & \textbf{G} &
\textbf{V} & \textbf{G} \\
\midrule

\makecell[l]{Baseline\\(no defense)} &
\makecell[c]{$ASR_{@once}$\\$ASR_{@max\_q}$} &
\makecell[c]{83.27\\100.00} &
\makecell[c]{92.73\\100.00} &
\makecell[c]{79.33\\93.00} &
\makecell[c]{88.00\\94.00} &
\makecell[c]{74.20\\94.00} &
\makecell[c]{82.40\\92.00} &
\makecell[c]{51.67\\100.00} &
\makecell[c]{86.33\\100.00} &
\makecell[c]{59.75\\91.00} &
\makecell[c]{63.25\\95.00} \\

PPL &
\makecell[c]{$ASR_{@once}$\\$ASR_{@max\_q}$} &
\makecell[c]{83.27\\100.00} &
\makecell[c]{92.73\\100.00} &
\makecell[c]{0.67\\\underline{1.00}} &
\makecell[c]{2.00\\2.00} &
\makecell[c]{77.80\\94.00} &
\makecell[c]{80.40\\92.00} &
\makecell[c]{51.67\\100.00} &
\makecell[c]{86.33\\100.00} &
\makecell[c]{64.00\\94.00} &
\makecell[c]{67.75\\96.00} \\

PG &
\makecell[c]{$ASR_{@once}$\\$ASR_{@max\_q}$} &
\makecell[c]{\textbf{0.00}\\\textbf{0.00}} &
\makecell[c]{\textbf{0.00}\\\textbf{0.00}} &
\makecell[c]{36.33\\60.00} &
\makecell[c]{59.33\\71.00} &
\makecell[c]{\textbf{0.00}\\\textbf{0.00}} &
\makecell[c]{\textbf{0.00}\\\textbf{0.00}} &
\makecell[c]{24.75\\57.00} &
\makecell[c]{47.83\\59.00} &
\makecell[c]{63.00\\94.00} &
\makecell[c]{66.50\\96.00} \\

WG &
\makecell[c]{$ASR_{@once}$\\$ASR_{@max\_q}$} &
\makecell[c]{\textbf{0.00}\\\textbf{0.00}} &
\makecell[c]{\textbf{0.00}\\\textbf{0.00}} &
\makecell[c]{2.00\\3.00} &
\makecell[c]{2.33\\3.00} &
\makecell[c]{\underline{1.00}\\\underline{2.00}} &
\makecell[c]{\underline{0.80}\\\underline{2.00}} &
\makecell[c]{6.83\\23.00} &
\makecell[c]{15.08\\26.00} &
\makecell[c]{25.00\\60.00} &
\makecell[c]{37.25\\72.00} \\

oss &
\makecell[c]{$ASR_{@once}$\\$ASR_{@max\_q}$} &
\makecell[c]{\underline{10.20}\\\underline{18.00}} &
\makecell[c]{\underline{13.53}\\\underline{17.00}} &
\makecell[c]{14.67\\18.00} &
\makecell[c]{15.33\\17.00} &
\makecell[c]{11.00\\17.00} &
\makecell[c]{12.20\\16.00} &
\makecell[c]{5.67\\26.00} &
\makecell[c]{14.75\\30.00} &
\makecell[c]{23.00\\62.00} &
\makecell[c]{28.75\\64.00} \\

$\{$PPL, PG$\}$ &
\makecell[c]{$ASR_{@once}$\\$ASR_{@max\_q}$} &
\makecell[c]{\textbf{0.00}\\\textbf{0.00}} &
\makecell[c]{\textbf{0.00}\\\textbf{0.00}} &
\makecell[c]{\underline{0.33}\\\underline{1.00}} &
\makecell[c]{1.00\\\underline{1.00}} &
\makecell[c]{\textbf{0.00}\\\textbf{0.00}} &
\makecell[c]{\textbf{0.00}\\\textbf{0.00}} &
\makecell[c]{24.75\\57.00} &
\makecell[c]{47.83\\59.00} &
\makecell[c]{42.75\\86.00} &
\makecell[c]{47.75\\87.00} \\

$\{$PPL, oss$\}$ &
\makecell[c]{$ASR_{@once}$\\$ASR_{@max\_q}$} &
\makecell[c]{\underline{10.20}\\\underline{18.00}} &
\makecell[c]{\underline{13.53}\\\underline{17.00}} &
\makecell[c]{0.67\\\underline{1.00}} &
\makecell[c]{1.00\\\underline{1.00}} &
\makecell[c]{11.80\\16.00} &
\makecell[c]{11.40\\15.00} &
\makecell[c]{5.67\\26.00} &
\makecell[c]{14.75\\30.00} &
\makecell[c]{15.25\\47.00} &
\makecell[c]{22.00\\55.00} \\

$\{$PG, WG$\}$ &
\makecell[c]{$ASR_{@once}$\\$ASR_{@max\_q}$} &
\makecell[c]{\textbf{0.00}\\\textbf{0.00}} &
\makecell[c]{\textbf{0.00}\\\textbf{0.00}} &
\makecell[c]{1.33\\3.00} &
\makecell[c]{2.33\\3.00} &
\makecell[c]{\textbf{0.00}\\\textbf{0.00}} &
\makecell[c]{\textbf{0.00}\\\textbf{0.00}} &
\makecell[c]{\underline{2.83}\\\underline{9.00}} &
\makecell[c]{\underline{6.42}\\\underline{10.00}} &
\makecell[c]{16.50\\43.00} &
\makecell[c]{25.75\\57.00} \\

$\{$PG, oss$\}$ &
\makecell[c]{$ASR_{@once}$\\$ASR_{@max\_q}$} &
\makecell[c]{\textbf{0.00}\\\textbf{0.00}} &
\makecell[c]{\textbf{0.00}\\\textbf{0.00}} &
\makecell[c]{9.33\\16.00} &
\makecell[c]{11.33\\14.00} &
\makecell[c]{\textbf{0.00}\\\textbf{0.00}} &
\makecell[c]{\textbf{0.00}\\\textbf{0.00}} &
\makecell[c]{4.83\\19.00} &
\makecell[c]{12.42\\21.00} &
\makecell[c]{14.75\\48.00} &
\makecell[c]{19.50\\50.00} \\

$\{$PPL, PG, WG$\}$ &
\makecell[c]{$ASR_{@once}$\\$ASR_{@max\_q}$} &
\makecell[c]{\textbf{0.00}\\\textbf{0.00}} &
\makecell[c]{\textbf{0.00}\\\textbf{0.00}} &
\makecell[c]{\textbf{0.00}\\\textbf{0.00}} &
\makecell[c]{\textbf{0.00}\\\textbf{0.00}} &
\makecell[c]{\textbf{0.00}\\\textbf{0.00}} &
\makecell[c]{\textbf{0.00}\\\textbf{0.00}} &
\makecell[c]{\underline{2.83}\\\underline{9.00}} &
\makecell[c]{\underline{6.42}\\\underline{10.00}} &
\makecell[c]{10.00\\30.00} &
\makecell[c]{19.00\\44.00} \\

$\{$PPL, PG, oss$\}$ &
\makecell[c]{$ASR_{@once}$\\$ASR_{@max\_q}$} &
\makecell[c]{\textbf{0.00}\\\textbf{0.00}} &
\makecell[c]{\textbf{0.00}\\\textbf{0.00}} &
\makecell[c]{\underline{0.33}\\\underline{1.00}} &
\makecell[c]{\underline{0.67}\\\underline{1.00}} &
\makecell[c]{\textbf{0.00}\\\textbf{0.00}} &
\makecell[c]{\textbf{0.00}\\\textbf{0.00}} &
\makecell[c]{4.83\\19.00} &
\makecell[c]{12.42\\21.00} &
\makecell[c]{10.25\\38.00} &
\makecell[c]{15.50\\42.00} \\

$\{$PG, WG, oss$\}$ &
\makecell[c]{$ASR_{@once}$\\$ASR_{@max\_q}$} &
\makecell[c]{\textbf{0.00}\\\textbf{0.00}} &
\makecell[c]{\textbf{0.00}\\\textbf{0.00}} &
\makecell[c]{\textbf{0.00}\\\textbf{0.00}} &
\makecell[c]{\textbf{0.00}\\\textbf{0.00}} &
\makecell[c]{\textbf{0.00}\\\textbf{0.00}} &
\makecell[c]{\textbf{0.00}\\\textbf{0.00}} &
\makecell[c]{\textbf{1.42}\\\textbf{6.00}} &
\makecell[c]{\textbf{3.83}\\\textbf{7.00}} &
\makecell[c]{\underline{4.50}\\\underline{16.00}} &
\makecell[c]{\underline{8.50}\\\underline{26.00}} \\

$\{$PPL, PG, WG, oss$\}$ &
\makecell[c]{$ASR_{@once}$\\$ASR_{@max\_q}$} &
\makecell[c]{\textbf{0.00}\\\textbf{0.00}} &
\makecell[c]{\textbf{0.00}\\\textbf{0.00}} &
\makecell[c]{\textbf{0.00}\\\textbf{0.00}} &
\makecell[c]{\textbf{0.00}\\\textbf{0.00}} &
\makecell[c]{\textbf{0.00}\\\textbf{0.00}} &
\makecell[c]{\textbf{0.00}\\\textbf{0.00}} &
\makecell[c]{\textbf{1.42}\\\textbf{6.00}} &
\makecell[c]{\textbf{3.83}\\\textbf{7.00}} &
\makecell[c]{\textbf{2.50}\\\textbf{10.00}} &
\makecell[c]{\textbf{7.25}\\\textbf{21.00}} \\

\bottomrule
\end{tabular*}
\captionsetup{
    justification=justified,
}
\caption{(P3) ASR results (\%) of the representative attacks against \textit{input filter} per-stage configurations. (V: Vicuna; G: GPT-3.5; Bold: stage-wise minimum; underlined: stage-wise second smallest values)}
\label{tab:appendix results 3.1}
\end{table*} 
\begin{table*}[t]
\centering
\footnotesize
\renewcommand{\arraystretch}{2}
\setlength{\tabcolsep}{2pt}
\newcommand{\headerpad}{\rule{0pt}{3.5ex}}
\begin{tabular*}{\textwidth}{@{\extracolsep{\fill}}lccccccccccc}
\toprule
\multirow{2}{*}{\textbf{\makecell[l]{Defense\\configuration}}} &
\multirow{2}{*}{\textbf{Metric}} &
\multicolumn{2}{c}{\textbf{Adaptive}} &
\multicolumn{2}{c}{\textbf{DSN}} &
\multicolumn{2}{c}{\textbf{Refusal}} &
\multicolumn{2}{c}{\textbf{SeqBreak}} &
\multicolumn{2}{c}{\textbf{ReNeLLM}} \\
\cmidrule(lr){3-4}
\cmidrule(lr){5-6}
\cmidrule(lr){7-8}
\cmidrule(lr){9-10}
\cmidrule(lr){11-12}
& &
\textbf{V} & \textbf{G} &
\textbf{V} & \textbf{G} &
\textbf{V} & \textbf{G} &
\textbf{V} & \textbf{G} &
\textbf{V} & \textbf{G} \\
\midrule

\makecell[l]{Baseline\\(no defense)} &
\makecell[c]{$ASR_{@once}$\\$ASR_{@max\_q}$} &
\makecell[c]{83.27\\\underline{100.00}} &
\makecell[c]{92.73\\100.00} &
\makecell[c]{79.33\\93.00} &
\makecell[c]{88.00\\94.00} &
\makecell[c]{74.20\\94.00} &
\makecell[c]{82.40\\92.00} &
\makecell[c]{51.67\\100.00} &
\makecell[c]{86.33\\100.00} &
\makecell[c]{59.75\\91.00} &
\makecell[c]{63.25\\95.00} \\

SR &
\makecell[c]{$ASR_{@once}$\\$ASR_{@max\_q}$} &
\makecell[c]{78.13\\\underline{100.00}} &
\makecell[c]{80.40\\96.00} &
\makecell[c]{35.00\\59.00} &
\makecell[c]{2.67\\3.00} &
\makecell[c]{54.80\\82.00} &
\makecell[c]{56.60\\69.00} &
\makecell[c]{31.33\\96.00} &
\makecell[c]{77.08\\\textbf{98.00}} &
\makecell[c]{60.25\\92.00} &
\makecell[c]{57.75\\94.00} \\

ICD &
\makecell[c]{$ASR_{@once}$\\$ASR_{@max\_q}$} &
\makecell[c]{67.80\\\underline{100.00}} &
\makecell[c]{86.67\\99.00} &
\makecell[c]{37.67\\68.00} &
\makecell[c]{\underline{0.67}\\\underline{1.00}} &
\makecell[c]{68.80\\98.00} &
\makecell[c]{69.80\\86.00} &
\makecell[c]{48.25\\99.00} &
\makecell[c]{90.00\\100.00} &
\makecell[c]{63.00\\97.00} &
\makecell[c]{66.25\\97.00} \\

RPO &
\makecell[c]{$ASR_{@once}$\\$ASR_{@max\_q}$} &
\makecell[c]{82.47\\\underline{100.00}} &
\makecell[c]{91.33\\100.00} &
\makecell[c]{70.33\\89.00} &
\makecell[c]{76.00\\82.00} &
\makecell[c]{61.60\\93.00} &
\makecell[c]{73.60\\86.00} &
\makecell[c]{41.25\\99.00} &
\makecell[c]{84.92\\100.00} &
\makecell[c]{61.75\\96.00} &
\makecell[c]{60.75\\93.00} \\

DPP &
\makecell[c]{$ASR_{@once}$\\$ASR_{@max\_q}$} &
\makecell[c]{81.40\\\underline{100.00}} &
\makecell[c]{92.67\\100.00} &
\makecell[c]{71.33\\93.00} &
\makecell[c]{66.00\\74.00} &
\makecell[c]{67.60\\92.00} &
\makecell[c]{80.40\\91.00} &
\makecell[c]{50.50\\99.00} &
\makecell[c]{88.75\\100.00} &
\makecell[c]{68.25\\95.00} &
\makecell[c]{70.25\\95.00} \\

SR $\Rightarrow$ DPP &
\makecell[c]{$ASR_{@once}$\\$ASR_{@max\_q}$} &
\makecell[c]{78.73\\\textbf{99.00}} &
\makecell[c]{72.47\\97.00} &
\makecell[c]{34.00\\57.00} &
\makecell[c]{2.67\\4.00} &
\makecell[c]{\underline{44.80}\\\underline{78.00}} &
\makecell[c]{55.40\\68.00} &
\makecell[c]{\textbf{27.33}\\\underline{95.00}} &
\makecell[c]{\underline{74.42}\\\textbf{98.00}} &
\makecell[c]{60.75\\\underline{90.00}} &
\makecell[c]{50.75\\87.00} \\

ICD $\Rightarrow$ SR &
\makecell[c]{$ASR_{@once}$\\$ASR_{@max\_q}$} &
\makecell[c]{\textbf{51.00}\\\underline{100.00}} &
\makecell[c]{\textbf{2.93}\\\textbf{22.00}} &
\makecell[c]{\textbf{17.00}\\\underline{39.00}} &
\makecell[c]{\textbf{0.00}\\\textbf{0.00}} &
\makecell[c]{56.80\\85.00} &
\makecell[c]{\underline{32.40}\\\underline{55.00}} &
\makecell[c]{45.00\\100.00} &
\makecell[c]{\textbf{66.50}\\\textbf{98.00}} &
\makecell[c]{59.25\\\underline{90.00}} &
\makecell[c]{\underline{43.00}\\\underline{80.00}} \\

SR $\Rightarrow$ ICD &
\makecell[c]{$ASR_{@once}$\\$ASR_{@max\_q}$} &
\makecell[c]{62.93\\\textbf{99.00}} &
\makecell[c]{\underline{37.93}\\\underline{78.00}} &
\makecell[c]{28.33\\54.00} &
\makecell[c]{\textbf{0.00}\\\textbf{0.00}} &
\makecell[c]{49.00\\82.00} &
\makecell[c]{42.80\\66.00} &
\makecell[c]{38.17\\97.00} &
\makecell[c]{76.50\\100.00} &
\makecell[c]{\underline{57.50}\\97.00} &
\makecell[c]{43.25\\\textbf{79.00}} \\

SR $\Rightarrow$ ICD $\Rightarrow$ DPP &
\makecell[c]{$ASR_{@once}$\\$ASR_{@max\_q}$} &
\makecell[c]{\underline{60.47}\\\textbf{99.00}} &
\makecell[c]{47.87\\81.00} &
\makecell[c]{\underline{19.00}\\\textbf{38.00}} &
\makecell[c]{2.00\\2.00} &
\makecell[c]{\textbf{29.60}\\\textbf{69.00}} &
\makecell[c]{\textbf{32.20}\\\textbf{51.00}} &
\makecell[c]{\underline{30.17}\\\textbf{90.00}} &
\makecell[c]{78.75\\\underline{99.00}} &
\makecell[c]{\textbf{48.50}\\\textbf{88.00}} &
\makecell[c]{\textbf{42.50}\\82.00} \\

\bottomrule
\end{tabular*}
\captionsetup{
    justification=justified,
}
\caption{(P3) ASR results (\%) of the representative attacks against \textit{input modification} per-stage configurations. (V: Vicuna; G: GPT-3.5; Bold: stage-wise minimum; underlined: stage-wise second smallest values)}
\label{tab:appendix results 3.2}
\end{table*} 
\begin{table*}[t]
\centering
\footnotesize
\renewcommand{\arraystretch}{2}
\setlength{\tabcolsep}{2pt}
\begin{tabular*}{\textwidth}{@{\extracolsep{\fill}}lccccccccccc}
\toprule
\multirow{2}{*}{\textbf{\makecell[l]{Defense\\configuration}}} &
\multirow{2}{*}{\textbf{Metric}} &
\multicolumn{2}{c}{\textbf{Adaptive}} &
\multicolumn{2}{c}{\textbf{DSN}} &
\multicolumn{2}{c}{\textbf{Refusal}} &
\multicolumn{2}{c}{\textbf{SeqBreak}} &
\multicolumn{2}{c}{\textbf{ReNeLLM}} \\
\cmidrule(lr){3-4}
\cmidrule(lr){5-6}
\cmidrule(lr){7-8}
\cmidrule(lr){9-10}
\cmidrule(lr){11-12}
& &
\textbf{V} & \textbf{G} &
\textbf{V} & \textbf{G} &
\textbf{V} & \textbf{G} &
\textbf{V} & \textbf{G} &
\textbf{V} & \textbf{G} \\
\midrule

\makecell[l]{Baseline\\(no defense)} &
\makecell[c]{$ASR_{@once}$\\$ASR_{@max\_q}$} &
\makecell[c]{83.27\\100.00} &
\makecell[c]{92.73\\100.00} &
\makecell[c]{79.33\\93.00} &
\makecell[c]{88.00\\94.00} &
\makecell[c]{74.20\\94.00} &
\makecell[c]{82.40\\92.00} &
\makecell[c]{51.67\\100.00} &
\makecell[c]{86.33\\100.00} &
\makecell[c]{59.75\\91.00} &
\makecell[c]{63.25\\95.00} \\

LG &
\makecell[c]{$ASR_{@once}$\\$ASR_{@max\_q}$} &
\makecell[c]{1.47\\6.00} &
\makecell[c]{1.40\\4.00} &
\makecell[c]{8.00\\13.00} &
\makecell[c]{11.33\\17.00} &
\makecell[c]{1.20\\4.00} &
\makecell[c]{2.00\\\underline{3.00}} &
\makecell[c]{35.25\\97.00} &
\makecell[c]{74.30\\97.00} &
\makecell[c]{16.00\\39.00} &
\makecell[c]{15.25\\41.00} \\

WG &
\makecell[c]{$ASR_{@once}$\\$ASR_{@max\_q}$} &
\makecell[c]{6.20\\24.00} &
\makecell[c]{5.07\\12.00} &
\makecell[c]{8.33\\14.00} &
\makecell[c]{12.00\\20.00} &
\makecell[c]{4.80\\9.00} &
\makecell[c]{3.20\\6.00} &
\makecell[c]{15.50\\65.00} &
\makecell[c]{30.33\\60.00} &
\makecell[c]{13.25\\38.00} &
\makecell[c]{12.25\\35.00} \\

XG &
\makecell[c]{$ASR_{@once}$\\$ASR_{@max\_q}$} &
\makecell[c]{35.73\\46.00} &
\makecell[c]{56.93\\96.00} &
\makecell[c]{62.33\\81.00} &
\makecell[c]{78.33\\92.00} &
\makecell[c]{40.20\\52.00} &
\makecell[c]{27.20\\54.00} &
\makecell[c]{51.17\\100.00} &
\makecell[c]{85.67\\100.00} &
\makecell[c]{56.75\\91.00} &
\makecell[c]{66.50\\96.00} \\

GR &
\makecell[c]{$ASR_{@once}$\\$ASR_{@max\_q}$} &
\makecell[c]{2.87\\12.00} &
\makecell[c]{3.40\\6.00} &
\makecell[c]{9.33\\14.00} &
\makecell[c]{10.67\\17.00} &
\makecell[c]{2.80\\6.00} &
\makecell[c]{2.20\\4.00} &
\makecell[c]{34.50\\90.00} &
\makecell[c]{60.25\\93.00} &
\makecell[c]{13.50\\42.00} &
\makecell[c]{14.50\\35.00} \\

$\{$LG, WG$\}$ &
\makecell[c]{$ASR_{@once}$\\$ASR_{@max\_q}$} &
\makecell[c]{0.73\\4.00} &
\makecell[c]{0.27\\\underline{3.00}} &
\makecell[c]{4.33\\7.00} &
\makecell[c]{6.67\\11.00} &
\makecell[c]{0.60\\2.00} &
\makecell[c]{1.80\\\textbf{2.00}} &
\makecell[c]{\underline{13.75}\\\underline{56.00}} &
\makecell[c]{27.00\\55.00} &
\makecell[c]{3.75\\11.00} &
\makecell[c]{\underline{2.50}\\9.00} \\

$\{$LG, XG$\}$ &
\makecell[c]{$ASR_{@once}$\\$ASR_{@max\_q}$} &
\makecell[c]{0.53\\4.00} &
\makecell[c]{1.07\\\underline{3.00}} &
\makecell[c]{5.67\\10.00} &
\makecell[c]{9.33\\15.00} &
\makecell[c]{\underline{0.20}\\\underline{1.00}} &
\makecell[c]{1.60\\\textbf{2.00}} &
\makecell[c]{35.25\\97.00} &
\makecell[c]{73.92\\97.00} &
\makecell[c]{10.25\\27.00} &
\makecell[c]{10.75\\33.00} \\

$\{$WG, XG$\}$ &
\makecell[c]{$ASR_{@once}$\\$ASR_{@max\_q}$} &
\makecell[c]{4.27\\15.00} &
\makecell[c]{4.40\\12.00} &
\makecell[c]{8.00\\14.00} &
\makecell[c]{11.67\\19.00} &
\makecell[c]{3.00\\7.00} &
\makecell[c]{2.80\\5.00} &
\makecell[c]{15.50\\65.00} &
\makecell[c]{27.08\\55.00} &
\makecell[c]{8.25\\25.00} &
\makecell[c]{9.50\\30.00} \\

$\{$WG, GR$\}$ &
\makecell[c]{$ASR_{@once}$\\$ASR_{@max\_q}$} &
\makecell[c]{2.67\\12.00} &
\makecell[c]{3.33\\5.00} &
\makecell[c]{6.33\\11.00} &
\makecell[c]{7.67\\11.00} &
\makecell[c]{3.00\\6.00} &
\makecell[c]{2.40\\5.00} &
\makecell[c]{14.83\\60.00} &
\makecell[c]{28.33\\56.00} &
\makecell[c]{1.75\\7.00} &
\makecell[c]{3.00\\9.00} \\

$\{$LG, WG, XG$\}$ &
\makecell[c]{$ASR_{@once}$\\$ASR_{@max\_q}$} &
\makecell[c]{\underline{0.40}\\\underline{2.00}} &
\makecell[c]{\underline{0.20}\\\underline{3.00}} &
\makecell[c]{3.67\\\underline{6.00}} &
\makecell[c]{5.67\\10.00} &
\makecell[c]{\textbf{0.00}\\\textbf{0.00}} &
\makecell[c]{1.60\\\textbf{2.00}} &
\makecell[c]{\underline{13.75}\\\underline{56.00}} &
\makecell[c]{25.83\\53.00} &
\makecell[c]{3.25\\10.00} &
\makecell[c]{\underline{2.50}\\9.00} \\

$\{$LG, WG, GR$\}$ &
\makecell[c]{$ASR_{@once}$\\$ASR_{@max\_q}$} &
\makecell[c]{\underline{0.40}\\3.00} &
\makecell[c]{\textbf{0.07}\\\textbf{1.00}} &
\makecell[c]{3.67\\7.00} &
\makecell[c]{\underline{4.67}\\\underline{7.00}} &
\makecell[c]{\underline{0.20}\\\underline{1.00}} &
\makecell[c]{\underline{1.40}\\\textbf{2.00}} &
\makecell[c]{\textbf{13.50}\\\textbf{53.00}} &
\makecell[c]{26.00\\\underline{51.00}} &
\makecell[c]{\underline{1.00}\\\underline{4.00}} &
\makecell[c]{\textbf{0.75}\\\textbf{2.00}} \\

$\{$LG, XG, GR$\}$ &
\makecell[c]{$ASR_{@once}$\\$ASR_{@max\_q}$} &
\makecell[c]{\textbf{0.27}\\\textbf{1.00}} &
\makecell[c]{\textbf{0.07}\\\textbf{1.00}} &
\makecell[c]{\underline{3.33}\\\underline{6.00}} &
\makecell[c]{5.00\\\underline{7.00}} &
\makecell[c]{\textbf{0.00}\\\textbf{0.00}} &
\makecell[c]{\textbf{1.20}\\\textbf{2.00}} &
\makecell[c]{28.67\\85.00} &
\makecell[c]{56.50\\89.00} &
\makecell[c]{2.50\\9.00} &
\makecell[c]{\underline{2.50}\\8.00} \\

$\{$WG, XG, GR$\}$ &
\makecell[c]{$ASR_{@once}$\\$ASR_{@max\_q}$} &
\makecell[c]{1.73\\7.00} &
\makecell[c]{3.20\\5.00} &
\makecell[c]{6.00\\10.00} &
\makecell[c]{7.33\\11.00} &
\makecell[c]{2.00\\5.00} &
\makecell[c]{2.00\\4.00} &
\makecell[c]{14.83\\60.00} &
\makecell[c]{\underline{25.58}\\\underline{51.00}} &
\makecell[c]{1.25\\5.00} &
\makecell[c]{\underline{2.50}\\\underline{7.00}} \\

$\{$LG, WG, XG, GR$\}$ &
\makecell[c]{$ASR_{@once}$\\$ASR_{@max\_q}$} &
\makecell[c]{\textbf{0.27}\\\textbf{1.00}} &
\makecell[c]{\textbf{0.07}\\\textbf{1.00}} &
\makecell[c]{\textbf{3.00}\\\textbf{5.00}} &
\makecell[c]{\textbf{4.33}\\\textbf{6.00}} &
\makecell[c]{\textbf{0.00}\\\textbf{0.00}} &
\makecell[c]{\textbf{1.20}\\\textbf{2.00}} &
\makecell[c]{\textbf{13.50}\\\textbf{53.00}} &
\makecell[c]{\textbf{24.92}\\\textbf{50.00}} &
\makecell[c]{\textbf{0.75}\\\textbf{3.00}} &
\makecell[c]{\textbf{0.75}\\\textbf{2.00}} \\

\bottomrule
\end{tabular*}
\captionsetup{
    justification=justified,
}
\caption{(P3) ASR results (\%) of the representative attacks against \textit{output guard} per-stage configurations. (V: Vicuna; G: GPT-3.5; Bold: stage-wise minimum; underlined: stage-wise second smallest values)}
\label{tab:appendix results 3.3}
\end{table*} 

\begin{table*}[t]
\centering
\footnotesize
\renewcommand{\arraystretch}{1.25}
\setlength{\tabcolsep}{2pt}
\newcommand{\headerpad}{\rule{0pt}{3.5ex}}
\begin{tabular*}{\textwidth}{@{\extracolsep{\fill}}lcccccccccc}
\toprule
\multirow{2}{*}{\headerpad\textbf{\makecell[l]{Cross-stage\\defense combination}}} 
& \multicolumn{2}{c}{\textbf{Adaptive}}
& \multicolumn{2}{c}{\textbf{DSN}}
& \multicolumn{2}{c}{\textbf{Refusal}}
& \multicolumn{2}{c}{\textbf{SeqBreak}}
& \multicolumn{2}{c}{\textbf{ReNeLLM}} \\
\cmidrule(lr){2-3}
\cmidrule(lr){4-5}
\cmidrule(lr){6-7}
\cmidrule(lr){8-9}
\cmidrule(lr){10-11}
& \textbf{V} & \textbf{G}
& \textbf{V} & \textbf{G}
& \textbf{V} & \textbf{G}
& \textbf{V} & \textbf{G}
& \textbf{V} & \textbf{G} \\
\midrule

Baseline (no defense)
& 100.00 & 100.00
& 93.00 & 94.00
& 94.00 & 92.00
& 100.00 & 100.00
& 91.00 & 95.00 \\

\midrule

\textbf{sec1} ($\{\text{PG},\text{WG},\text{oss}\}$)
& 0.00 & 0.00
& 0.00 & 0.00
& 0.00 & 0.00
& 6.00 & 7.00
& 16.00 & 26.00 \\

\textbf{util1} ($\{\text{PG},\text{oss}\}$)
& 0.00 & 0.00
& 16.00 & 14.00
& 0.00 & 0.00
& 19.00 & 21.00
& 48.00 & 50.00 \\

\textbf{mem1} ($\{\text{PG},\text{WG}\}$)
& 0.00 & 0.00
& 3.00 & 3.00
& 0.00 & 0.00
& 9.00 & 10.00
& 43.00 & 57.00 \\

\textbf{sec2} (ICD $\Rightarrow$ SR)
& 100.00 & 22.00
& 39.00 & 0.00
& 85.00 & 55.00
& 100.00 & 98.00
& 90.00 & 80.00 \\

\textbf{util2} (SR)
& 100.00 & 96.00
& 59.00 & 3.00
& 82.00 & 69.00
& 96.00 & 98.00
& 92.00 & 94.00 \\

\textbf{sec3} ($\{\text{LG},\text{WG},\text{GR}\}$)
& 3.00 & 1.00
& 7.00 & 7.00
& 1.00 & 2.00
& 53.00 & 51.00
& 4.00 & 2.00 \\

\textbf{util3} ($\{\text{LG},\text{WG}\}$)
& 4.00 & 3.00
& 7.00 & 11.00
& 2.00 & 2.00
& 56.00 & 55.00
& 11.00 & 9.00 \\

\midrule

\textbf{util1} $\Rightarrow$ \textbf{util3}
& 3.00 & 3.00
& 0.00 & 0.00
& 2.00 & 1.00
& 20.00 & 16.00
& 4.00 & 4.00 \\

\textbf{util1} $\Rightarrow$ \textbf{util2}
& 0.00 & 0.00
& 9.00 & 2.00
& 0.00 & 0.00
& 17.00 & 16.00
& 42.00 & 50.00 \\

\textbf{util1} $\Rightarrow$ \textbf{sec3}
& 2.00 & 1.00
& 0.00 & 0.00
& 1.00 & 1.00
& 19.00 & 15.00
& 1.00 & 1.00 \\

\textbf{util2} $\Rightarrow$ \textbf{util3}
& 10.00 & 4.00
& 6.00 & 2.00
& 5.00 & 3.00
& 56.00 & 48.00
& 11.00 & 15.00 \\

\midrule

\textbf{util1} $\Rightarrow$ \textbf{util2} $\Rightarrow$ \textbf{sec3}
& 0.00 & 0.00
& 1.00 & 1.00
& 0.00 & 0.00
& 5.00 & 6.00
& 5.00 & 7.00 \\

\textbf{util1} $\Rightarrow$ \textbf{util2} $\Rightarrow$ \textbf{util3}
& 0.00 & 0.00
& 2.00 & 0.00
& 0.00 & 0.00
& 9.00 & 6.00
& 7.00 & 8.00 \\

\textbf{mem1} $\Rightarrow$ \textbf{util2} $\Rightarrow$ \textbf{sec3}
& 0.00 & 0.00
& 0.00 & 0.00
& 0.00 & 0.00
& 6.00 & 7.00
& 2.00 & 3.00 \\

\textbf{sec1} $\Rightarrow$ \textbf{util2} $\Rightarrow$ \textbf{util3}
& 0.00 & 0.00
& 1.00 & 0.00
& 0.00 & 0.00
& 4.00 & 6.00
& 4.00 & 5.00 \\

\bottomrule
\end{tabular*}
\captionsetup{
    justification=justified,
}
\caption{(P4) $ASR_{@max\_q}$ results (\%) of the representative attacks against cross-stage defense combinations. (V: Vicuna; G: GPT-3.5)}
\label{tab:appendix results 4.1}
\end{table*} 
\begin{table*}[t]
\centering
\footnotesize
\renewcommand{\arraystretch}{1.25}
\begin{tabular}{lcccc}
\toprule
\makecell[l]{\textbf{Cross-stage defense}\\\textbf{combination}} &
$\overline{\%{\downarrow}ASR}$ &
$\overline{\%{\uparrow}U}$ &
\makecell[c]{$\resultParam$ \textbf{(B)}} &
\makecell[c]{$\resultT$ \textbf{(s)}} \\

\midrule
\textbf{sec1} ($\{\text{PG},\text{WG},\text{oss}\}$) & 94.2 & -9.9 & 27.1 & 0.27 \\
\textbf{util1} ($\{\text{PG},\text{oss}\}$) & 82.3 & -7.0 & 20.1 & 0.23 \\
\textbf{mem1} ($\{\text{PG},\text{WG}\}$) & 86.7 & -9.6 & 7.1 & -0.02 \\
\cmidrule{1-1}
\textbf{sec2} (ICD $\Rightarrow$ SR) & 30.5 & -13.9 & 0.0 & -0.88 \\
\textbf{util2} (SR) & 18.1 & -6.7 & 0.0 & -0.51 \\
\cmidrule{1-1}

\textbf{sec3} ($\{\text{LG},\text{WG},\text{GR}\}$) & 86.7 & -8.9 & 23.0 & 1.85 \\
\textbf{util3} ($\{\text{LG},\text{WG}\}$) & 83.7 & -7.4 & 15.0 & 0.13 \\

\midrule
\textbf{util1} $\Rightarrow$ \textbf{util3} & 94.6 & -7.8 & 35.1 & 0.48\\
\textbf{util1} $\Rightarrow$ \textbf{util2} & 85.6 & -8.1 & 20.1 & -0.27 \\
\textbf{util1} $\Rightarrow$ \textbf{sec3} & 95.9 & -8.2 & 43.1 & 2.60 \\
\textbf{util2} $\Rightarrow$ \textbf{util3} & 83.7 & -9.2 & 15.0 & -0.35 \\

\midrule
\textbf{util1} $\Rightarrow$ \textbf{util2} $\Rightarrow$ \textbf{sec3} & 97.4 & -9.2 & 43.1 & 2.16 \\
\textbf{util1} $\Rightarrow$ \textbf{util2} $\Rightarrow$ \textbf{util3} & 96.7 & -10.2 & 35.1 & -0.03 \\
\textbf{mem1} $\Rightarrow$ \textbf{util2} $\Rightarrow$ \textbf{sec3} & 98.2 & -14.3 & 30.1 & 1.97 \\
\textbf{sec1} $\Rightarrow$ \textbf{util2} $\Rightarrow$ \textbf{util3} & 97.9 & -14.9 & 42.1 & 0.09 \\

\bottomrule
\end{tabular}

\caption{(P4) Security, utility and efficiency results of cross-stage defense combinations.}
\label{tab:appendix results 4.2}
\end{table*} 
\begin{table*}[t]
\centering
\footnotesize
\renewcommand{\arraystretch}{2}
\setlength{\tabcolsep}{1pt}
\newcommand{\headerpad}{\rule{0pt}{3.5ex}}
\begin{tabular*}{\textwidth}{@{\extracolsep{\fill}}lccccccccc}
\toprule
\headerpad\textbf{Defense}
& \textbf{Adaptive}
& \textbf{DSN}
& \textbf{Refusal}
& \textbf{SeqBreak}
& \textbf{ReNeLLM}
& $\overline{\%{\downarrow}ASR}$
& $Utility$
& $\resultU$
& $FPR$ \\
\midrule

\multicolumn{10}{c}{GPT-5.4-mini} \\

\midrule
\makecell[l]{Baseline\\(no defense)}
& 74.00 & 2.00 & 5.00 & 7.50 & 52.00 & 0.0 & 84.2 & 0.0 & 0.0 \\

\makecell[l]{
\textbf{util1} $\Rightarrow$ \textbf{util2} $\Rightarrow$ \textbf{sec3} \\
($\{\text{PG},\hspace{-0.2em}\text{OSS}\}\hspace{-0.2em}\Rightarrow\hspace{-0.2em}
[\text{SR}]\hspace{-0.2em}\Rightarrow\hspace{-0.2em}
\{\text{LG},\hspace{-0.2em}\text{WG},\hspace{-0.2em}\text{GR}\}$)
}
& 0.00 & 0.00 & 0.00 & 0.00 & 0.00 & 100.0 & 85.4 & 1.4 & 7.6 \\

\makecell[l]{
\textbf{util1} $\Rightarrow$ \textbf{util3} \\
($\{\text{PG},\hspace{-0.2em}\text{OSS}\}\hspace{-0.2em}\Rightarrow\hspace{-0.2em}
\{\text{LG},\hspace{-0.2em}\text{WG}\}$)
}
& 0.00 & 0.50 & 0.00 & 1.00 & 0.00 & 92.3 & 82.6 & -1.9 & 8.0 \\

\makecell[l]{
\textbf{mem1} \\
($\{\text{PG},\hspace{-0.2em}\text{WG}\}$)
}
& 0.00 & 0.00 & 0.00 & 0.50 & 21.50 & 90.4 & 78.9 & -6.3 & 1.2 \\

\midrule
\multicolumn{10}{c}{gemini-3.1-flash-lite} \\

\midrule
\makecell[l]{Baseline\\(no defense)}
& 96.00 & 28.50 & 66.50 & 99.50 & 92.00 & 0.0 & 81.2 & 0.0 & 0.0 \\

\makecell[l]{
\textbf{util1} $\Rightarrow$ \textbf{util2} $\Rightarrow$ \textbf{sec3} \\
($\{\text{PG},\hspace{-0.2em}\text{OSS}\}\hspace{-0.2em}\Rightarrow\hspace{-0.2em}
[\text{SR}]\hspace{-0.2em}\Rightarrow\hspace{-0.2em}
\{\text{LG},\hspace{-0.2em}\text{WG},\hspace{-0.2em}\text{GR}\}$)
}
& 0.00 & 0.00 & 0.00 & 16.50 & 0.00 & 96.3 & 80.8 & -0.5 & 8.0 \\

\makecell[l]{
\textbf{util1} $\Rightarrow$ \textbf{util3} \\
($\{\text{PG},\hspace{-0.2em}\text{OSS}\}\hspace{-0.2em}\Rightarrow\hspace{-0.2em}
\{\text{LG},\hspace{-0.2em}\text{WG}\}$)
}
& 0.00 & 0.00 & 0.00 & 13.00 & 6.00 & 96.1 & 79.2 & -2.5 & 7.2 \\

\makecell[l]{
\textbf{mem1} \\
($\{\text{PG},\hspace{-0.2em}\text{WG}\}$)
}
& 0.00 & 0.00 & 0.00 & 14.00 & 57.00 & 84.8 & 73.0 & -10.1 & 1.2 \\

\bottomrule
\end{tabular*}
\captionsetup{
    justification=justified,
}
\caption{ASR (\%), Utility and FPR results of generalizability experiment on GPT-5.4-mini, gemini-3.1-flash-lite, HarmBench and XSTest.}
\label{tab:appendix results 5.1}
\end{table*} 

\end{document}